# A 4D-STEM Tomographic Framework Assisted by Object Tracking for Nanoparticle Structure Determination

*Saleh Gholam*, Arno Annys, Irina Skvortsova, Erica Cordero Oyonarte, Amirhossein Hajizadeh, Philippe Boullay, Johan Hofkens, Johan Verbeeck, Joke Hadermann**

S. G., A. A., I. S., A. H., J. V., J. Ha.
EMAT, Department of Physics, University of Antwerp, Groenenborgerlaan 171, Antwerp 2020, Belgium
E-mails: saleh.gholam@gmail.com; joke.hadermann@uantwerpen.be

E. C. O., P. B.
CRISMAT, Normandie Université, ENSICAEN, UNICAEN, CNRS UMR 6508, 6 Bd Maréchal Juin, F-14050 Cedex Caen, France

I. S., J. Ho.
Department of Chemistry, KU Leuven, Leuven 3001, Belgium

**Funding**

1. Research Foundation Flanders (FWO, Belgium) projects SBO S000121N (AutomatED), SBO 1SHA024N and G069925N.
2. The European Union’s Horizon 2020 research and innovation programme under the Marie-Sklodowska-Curie grant agreement No 956099 (NanED – Electron Nanocrystallography – H2020-MSCA-ITN).
3. Research Fund of the University of Antwerp through BOF TOP 38689 and BOF/SEP Project 53221.
4. This study was also funded by the European Union by ERC grant (REACT, 101199099). Views and opinions expressed are however those of the author(s) only and do not necessarily reflect those of the European Union or the European Research Council Executive Agency. Neither the European Union nor the granting authority can be held responsible for them.
5. The European Union's Horizon Europe programme under grant agreement no. 101094299 (Impress). Views and opinions expressed are however those of the authors only and do not necessarily reflect those of the European Union or the European Research Executive Agency (REA). Neither the European Union nor the granting authority can be held responsible for them.

**Abstract**

Three-dimensional electron diffraction (3D ED) has emerged as a powerful method for solving the structures of sub-micron-sized particles down to nanoparticles. However, it faces technical challenges when applied to beam-sensitive samples or agglomerated nanoparticles. This study presents a novel approach that combines 4D-STEM tomography with object tracking and segmentation algorithms to overcome these limitations and achieve single-crystalline 3D ED datasets from nanopowder samples. The method and data quality are assessed on brookite $TiO_2$ nanorods and beam-sensitive $CsPbBr_3$ nanoparticles. To finely sample the reciprocal-space, the data acquisition was automated to acquire hundreds of 4D-STEM scans at fine tilt steps using a slightly convergent beam (0.6 to 1 mrad). Compared to conventional 3D ED methodologies, the proposed method provides enhanced signal-to-noise ratio, low illumination time for reducing beam damage, and the ability to analyze multiple particles from a single tomographic dataset. The procedure is optimized to be feasible using commercially available desktops and detectors. This extends the method applicability to systems and samples that were very challenging for conventional 3D ED methods, by eliminating several technical challenges for the data acquisition.

## 1. Introduction

3D electron diffraction (3D ED) has proven to be a powerful method for solving the structure of nanoparticles with sub-angstrom accuracy.[1, 2] Similar to single-crystal X-ray diffraction, 3D ED relies on diffraction tomography; however, it is typically performed in a transmission electron microscope (TEM). In this technique, diffraction patterns are collected from a single-crystalline nanoparticle in a quasi-parallel regime, while the TEM stage is sequentially or continuously tilting. Thanks to the strong interaction of the accelerated electrons with matter, electron diffraction allows the study of particles as small as ten nanometers.[3] However, a successful structure solution requires diffraction patterns from a single-crystalline region. If the sample consists of aggregates or multi-domain particles, obtaining a 3D ED dataset from a single, coherent lattice becomes challenging. The overlap of diffraction patterns from different lattices complicates the analysis and introduces errors on the reflection intensities preventing a reliable structure solution and refinement.

Even when suitable crystals are present, data collection poses several technical challenges. For instance, TEM stage goniometers typically suffer from mechanical imperfections during stage tilt. Consequently, the particle can easily move outside the pre-defined illuminated area during tomography. If the sample has low-contrast or is beam-sensitive, this aggravates the data collection challenges. These conditions can lead to low-quality 3D ED datasets having low reciprocal-space completeness, signs of structural damage, limited diffraction resolution or poor signal-to-noise ratio.

Several methods have been proposed in the literature to overcome such technical difficulties. To minimize the chance of illuminating the surrounding particles, the beam size is set at a minimum with respect to the region of interest. This also enhances the signal strength by focusing the beam flux on the particle and minimizing the background signal coming from the TEM grid support.[4] This is particularly of importance for small nanoparticles with low scattering power. To facilitate locating low-contrast particles, it is possible to switch from the TEM mode to the scanning mode (STEM) with a focused probe, and benefit from the high-angle annular dark-field (HAADF) detector. In this condition, the probe is afterwards re-opened to achieve a quasi-parallel beam for diffraction pattern collection.[5] Nevertheless, the success of all these approaches depends on an effective particle tracking strategy to tackle the movement of the particle during tomography.

In some 3D ED methodologies, it is possible to iteratively pause the process, and manually re-center the particle;[5] nonetheless, these methods are time-consuming and susceptible to operational errors. Moreover, such an approach is not aligned with the increasing adoption of higher throughput 3D ED methodologies, such as continuous rotation electron diffraction (CRED). This method has effectively gained terrain on precession-assisted 3D ED,[6, 7] thanks to its hardware and software simplicity as well as thorough sampling of the reciprocal space.[8-10] This has facilitated the development of automated particle tracking approaches to achieve higher throughput and minimize the exposure dose for beam-sensitive samples.[4, 9-13] These methods typically exhibit a high success rate for well-isolated single-crystals; however, they are severely challenged when facing particle clusters or agglomerations. Moreover, they typically require additional calibration procedures, such as calibrating beam shift and stage tilt speed, and they rely on the reproducibility of the stage movements during consecutive tilts in the case of pre-calibrated approaches.[13] Notably, since these methods are applied during the data acquisition at the microscope, any unpredicted error would either necessitate repeating the data acquisition by the user or directly translate to the loss of data in high-throughput methods.

In a wider perspective, most of the challenges on the particle tracking in CRED arise from the lack of simultaneous access to real-space information during diffraction tomography. Such duality is unified in 4-dimensional scanning transmission electron microscopy (4D-STEM).[14] In this method, a quasi-parallel probe scans a region, and the resulting diffraction patterns are collected at each probe position using a fast pixelated detector.[15] During post-processing, various virtual images can be calculated using the collected diffraction patterns to achieve real space information.[16] By combining 4D-STEM with tomography, a 3D ED dataset can be digitally extracted from a region of interest (ROI) during the post-processing. This was first pioneered by Gallagher-Jones et al. for protein structural analysis, under the label of NanoEDT.[17] Similarly, Saha et al. managed to solve the structure of a metal-organic framework with this method, albeit at substantially faster scan rates and on sub-micron-sized particles.[18] However, a universal and systematic method for tracking and defining ROIs is lacking in this approach. Additionally, in both these studies, large static tilt steps (1°) were employed which poorly samples the reflection intensity profiles, also known as rocking curves.[19] Furthermore, the datasets were as large as several hundreds of GB to a few TB, complicating data management and analysis.

A similar approach, Scanning Precession Electron Diffraction Tomography (SPET) [16, 20], provides better integration of the reflection intensities thanks to the adoption of the precession. This was recently employed for the accurate structural analysis of thin-films along their width.

[16, 20] Particle tracking was not needed in this work due to the geometry of the films.[21, 22] The precession is, however, costly and complicates the TEM setup. It also typically leads to higher 4D-STEM acquisition time, since the TEM deflection coils usually constrain the precession frequency. This can lead to longer data acquisition time, increasing the chances for contamination growth and beam damage.

In this study, we propose a 4D-STEM tomography approach employing a slightly convergent probe and fine tilt increments to enable effective sampling of reciprocal space, and thus subsequent dynamical refinement to tackle the discrepancies in structure refinement due to multiple scattering. By automating the acquisition workflow, hundreds of fast 4D-STEM scans are collected using an event-based direct electron detector.[23, 24] This allows the application of object tracking and segmentation algorithms to the resulting 4D-STEM tomograms for robust and versatile extraction of 3D ED datasets from selected regions. Owing to its wide field of view in the scans and nanometric spatial resolution, this approach enables the extraction of multiple 3D ED datasets from distinct regions within a single tomogram. We demonstrate the capabilities of the method on challenging systems, including packed particles in close proximity, agglomerates, and beam-sensitive halide perovskites with particle sizes down to 30 nm. The proposed approach also facilitates structural analysis of custom nanometer-scale regions of particles. The approach can be adopted for study a broad range of materials, a task which can impose challenges on conventional 3D ED techniques.

## 2. Results

### 2.1. 3D ED Reconstruction Workflow

To solve the structure of nanoparticles using 3D ED, we used a new strategy that combines 4D-STEM tomography with automated object tracking in the post-processing phase. The proposed workflow can be divided into five different steps, as illustrated in **Figure 1:** 1) acquisition of 4D-STEM scans in a tilt series, 2) calculation of virtual images for visualizing the scanned area, 3) object tracking and segmentation of a ROI based on these virtual images, 4) extraction of 3D ED frames from the segmented regions, and 5) data reduction and structure solution.

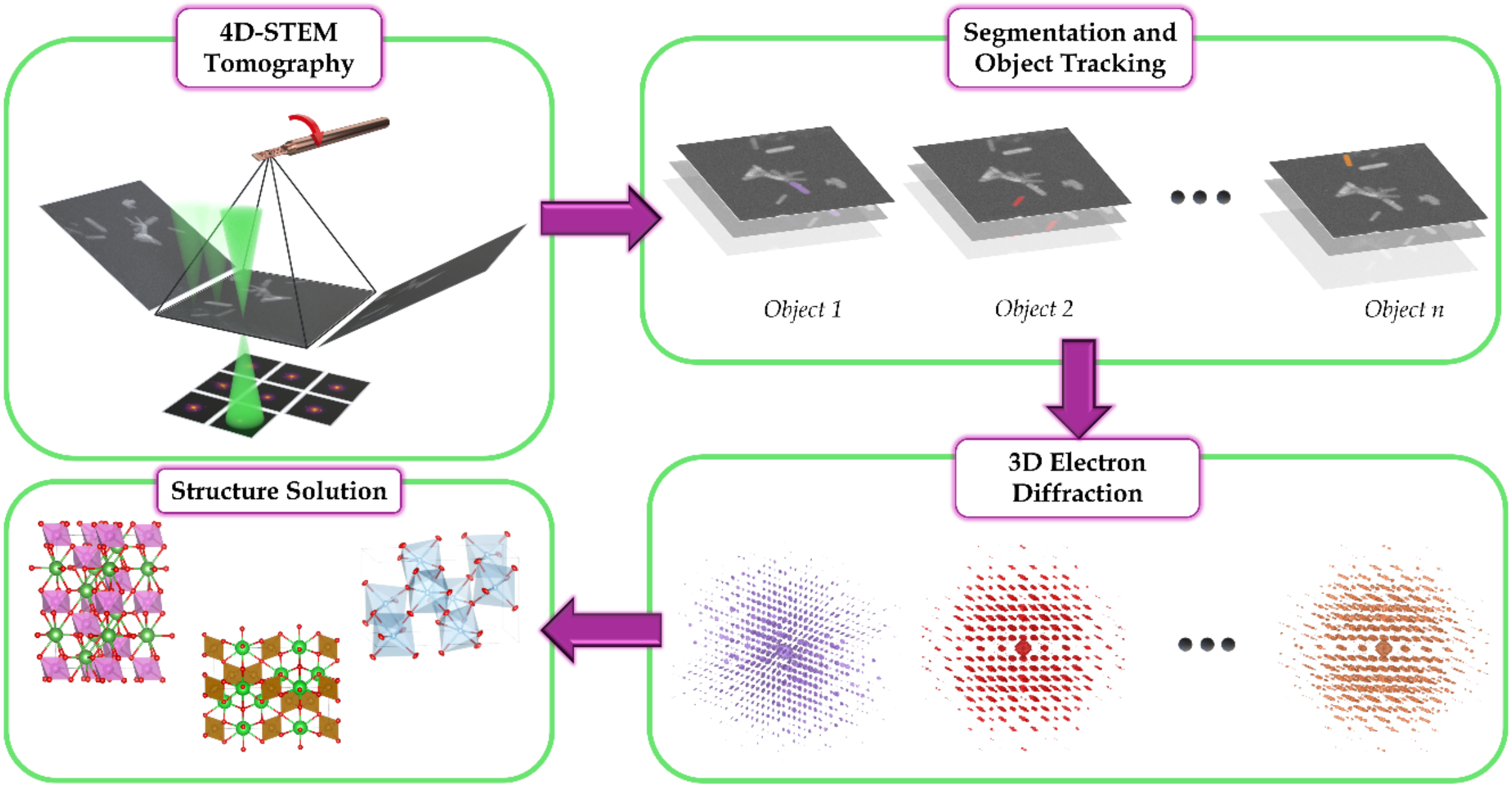

**Figure 1**. Schematic illustration of the data acquisition and analysis using 4D-STEM tomography for structure solution.

To achieve an ideal quality in 3D ED, thorough sampling of reciprocal space is essential. This allows the assignment of integrated intensities to reflections, a significant advantage for structure solution and refinement in the kinematical approximation, and an indispensable requirement for accurate structure analysis by taking into account dynamical scattering effects.[8, 25] Such sampling is typically achieved either by precession-assisted 3D ED in step-wise data collection[26] or by continuous rotation.[8, 27] In this study, such sampling quality was achieved using fine tilt steps, that is, 0.1° to 0.25°, as well as slight convergence of the beam to partially cover the information between the steps. Therefore, hundreds of 4D-STEM scans were performed using a commercially available, fast, event-based direct electron detector and by automating the data acquisition process. In addition to these measures, concerns such as beam damage and contamination growth were minimized by employing a minimal fluence. As a result, sparse diffraction patterns were collected at individual scan points (red in **Figure 2**), which could be summed over a region to achieve suitable patterns with distinct reflections and high reciprocal-space resolution for structure solution purposes (blue in **Figure 2**).

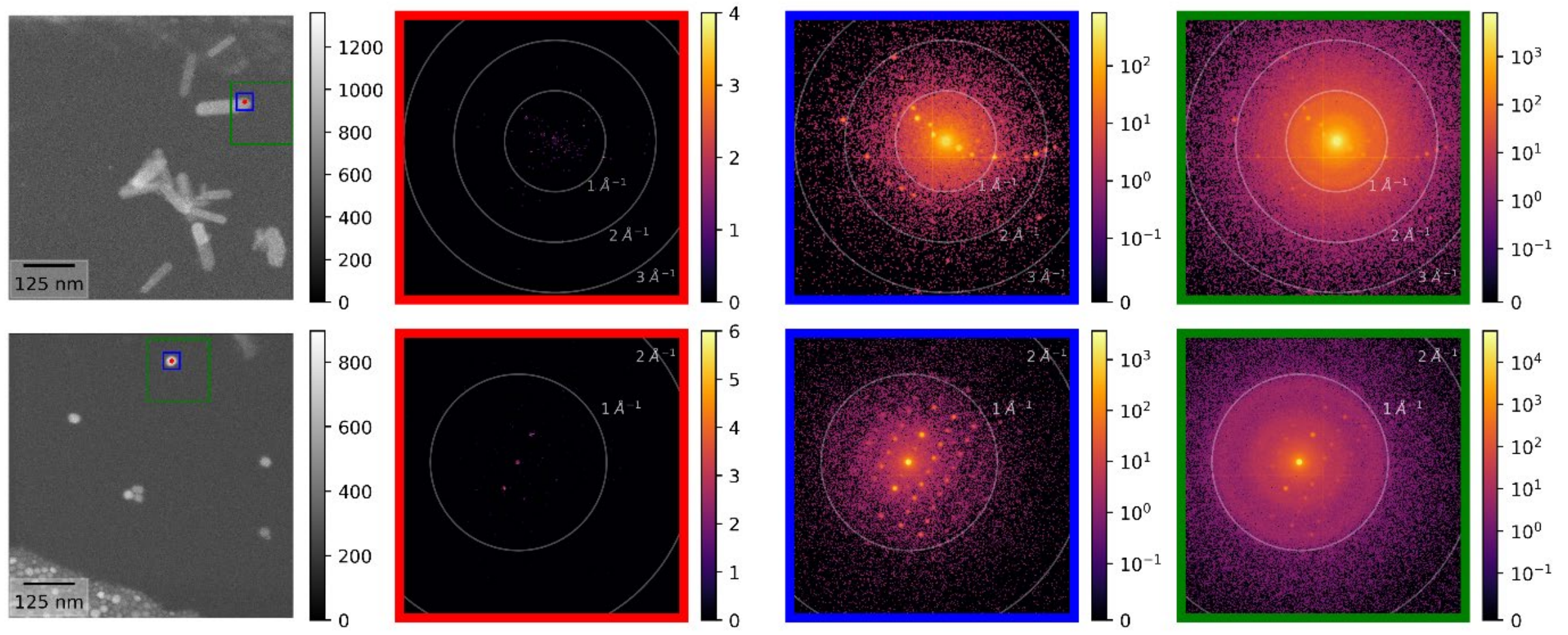

**Figure 2**. Diffraction pattern of a single point versus a small or large region in tomograms DS-1 (top row) and DS-3 (bottom row). The regions and their diffraction patterns share the same border colors.

In this methodology, particle tracking is postponed to the post-processing stage. This was performed on a set of virtual images calculated by summing the total number of detected electrons at each probe position. Owing to the slight convergence of the beam, these images have nanometric resolution. Next, object-tracking algorithms were applied to a user-defined region within these images, and a 3D ED dataset was extracted by summing the diffraction patterns of the tracked region. Owing to the wide scanned area, several datasets can be extracted from a single 4D-STEM tomogram.

In this study, two algorithms were used for tracking: the Discriminative Correlation Filter with Channel and Spatial Reliability (CSRT) [28] implemented in the OpenCV suite [29] and Segment-Anything-Model 2 (SAM2) [30] developed by Meta AI. Overall, the contrast of the image and the edges of the particles play significant roles in the success of the tracking procedure. Therefore, the navigation images can be fine-tuned to possess higher contrast to facilitate tracking. For instance, the calculation of virtual annular dark field images can be of interest. In this study, since the direct beam was almost fully saturated on the detector, the images calculated on the whole detector area resemble virtual annular dark field and no further tuning was required. While SAM2 itself applies a segmentation process on the object, this was added as an extra step to the CSRT results by Gaussian blurring and binarization of the tracked region (**Figure S4**). The segmentation step ensured that only diffraction patterns from the particle or the area of interest contributed to the final 3D ED frames. This effectively increases the signal-to-noise ratio by excluding contributions from the supporting membrane, as illustrated by the blue versus green regions in **Figure 2**. It should be noted that any segmentation and tracking algorithm can be employed in this methodology. Moreover, the resulting segmented regions can be partitioned further to provide a sub-particle structure. In this regard, we study the case of particle edges in Section 2.2.3.

## 2.2. Structure Solution and Refinement

Two samples were studied in this work: $TiO_2$ brookite nanorods and $CsPbBr_3$ nanoparticles. The challenges in performing conventional 3D ED methods differ for these two samples. Although 3D ED can be performed on the brookite sample[3], it is cumbersome because the nanorods form agglomerates. These agglomerates sometimes have strong bonds, making them very challenging to be dispersed. The $CsPbBr_3$ sample introduces extra challenges for 3D ED due to its 30-nm particle size and beam-sensitivity.[31] The beam sensitivity necessitates employing a low dose condition, in which case the $CsPbBr_3$ particles exhibit low contrast in TEM making it difficult to locate them. Additionally, contamination can rapidly grow in the illuminated area because of the ligands present in the solvent to stabilize the nanoparticles during synthesis. In total, three 4D-STEM tomograms are discussed in this study, as listed in **Table 1**. 3D ED analyses are mainly performed on DS-1 and DS-3 datasets, and DS-2 dataset is used to analyze the effect of the tilt step on the tomogram quality as the large number of 4D-STEM scans significantly increases the calculation time.

**Table 1**. Experimental details of the acquired 4D-STEM tomograms.

| Dataset | **DS-1** | **DS-2** | **DS-3** |
|---|---|---|---|
| Sample | $TiO_2$ | $TiO_2$ | $CsPbBr_3$ |
| Tilt Range [°] | 80 | 100 | 70 |
| Tilt Step [°] | 0.20 | 0.10 | 0.25 |
| FOV [$\mu m^2$] | 0.48 | 4.32 | 0.48 |
| Dwell Time [µs] | 20 | 25 | 50 |
| Acquisition Time [min] [a] | 40 | 125 | 65 |
| C2 Aperture [µm] | 20 | 20 | 10 |
| Convergence Semiangle [mrad] | 1.2 | 1.2 | 0.6 |
| Probe FWHM [nm] **[b]** | 2.1 | 2.1 | 3.9 |
| **Fluence** per Tilt [$e^- Å^{-2}$] | 1.11 ± 0.06 | 0.15 ± 0.01 | 1.40 ± 0.15 |
| Total **Fluence** [$e^- Å^{-2}$] | 444 ± 24 | 153 ± 8 | 393 ± 43 |

[a)] In addition to the scanning time, the acquisition time includes all other adjustments during the experiment, such as stage tilt.
[b)] The probe FWHM was calculated based on the image of the probe reported in **Figure S3**.

Several 3D ED datasets were extracted from the acquired tomograms, and the data reduction process was performed using the PETS2 software.[6] During this process, the diffracted intensities are located on each 2D frame, the reciprocal space is reconstructed in 3D, the unit cell of the particle is found and then, the intensities of the reflections are integrated over their intensity profile (rocking curve). At the end, a list of reflections with their intensities and the corresponding uncertainties are reached, which is used for the structure solution process. The quality of the datasets and the integration process was evaluated based on the R-factors at the end of the data reduction process. These R-factors are the internal consistency among symmetry-equivalent reflections ($R_{int}$) and redundancy-independent $R_{int}$ ($R_{meas}$). The details of all R-factors used in this section can be found in the Supporting Information. After the initial structure solution, the resulting structures were evaluated based on the refinement R-factor between the observed and calculated structure factors ($R_{obs}$). This was conducted at 5 stages: (1) after kinematical refinement ($R_{obs}^{kin}$); (2) after introducing extinction correction to the kinematically refined structure ($R_{obs}^{ext}$); (3) after dynamical refinement on the kinematically refined structure at stage 1 ($R_{obs}^{dyn}$); (4) after culling poorly matched reflections from the dynamical refinement ($R_{obs}^{cull}$); and (5) after introducing anisotropic atomic displacements parameters (ADPs) to the structure at stage 4 ($R_{obs}^{aniso}$).

During the dynamical refinement stages, the integrated intensities are refined per frame to take multiple scattering into account via Bloch wave formalism.[25] This in contrast to kinematical refinement in which the structure refinement is performed on structure factors based on intensities resulting from merging the same reflection across the different frames. The dynamical approach allows refining atomic movements in the structure due to the thermal motion with higher certainties. The culling of the reflections in stage 4 aimed to discard only reflections with saturated intensities, to decrease the uncertainties on the structure caused purely by the detector during the kinematical refinement, or reflections that were poorly fitted during data reduction. The selection of the culling threshold is discussed in the Supporting Information.

### *2.2.1. $TiO_2$ Brookite Nanorods*

SAM2 was used for segmentation and tracking on DS-1 taken from a sample comprising brookite particles. **Figure 3** depicts the tracking results for five different particles, demonstrated at 5° tilt intervals, and **Figure 4** demonstrates the corresponding diffraction patterns for T-P1 ($TiO_2$-Particle 1) in this tilt series. Even though the fluence was merely 1.11 ± 0.06 $e^{-}Å^{-2}$, in each 4D-STEM scan, the resolution of the diffraction patterns reached 3 $Å^{-1}$, which is excellent for structure solution purposes. During the data reduction process, in all extracted 3D ED datasets, except for T-P5, the unit cell of the brookite particles was successfully determined without any signs of other domains or phases. The acquisition of such datasets can be highly challenging for conventional 3D ED methods, as the distance between T-P1 and T-P2 was merely 20–30 nm in the whole tilt range, and T-P3 was part of an agglomerate. For T-P5, two different domains were found in the extracted 3D ED dataset. This could also be observed in some virtual images as a different contrast at the bottom and top of the particle (**Figure 2**). In this dataset, both domains exhibited the brookite unit cell, each indexing approximately 40% of the peaks in the 3D ED datasets. Using the information obtained during the 3D ED analysis, these domains could be visualized by virtual detectors on the related 4D-STEM scans (**Figure S5**). This dataset was excluded from further analysis because of the presence of different domains with overlapping reflections, which was beyond the scope of the current study.

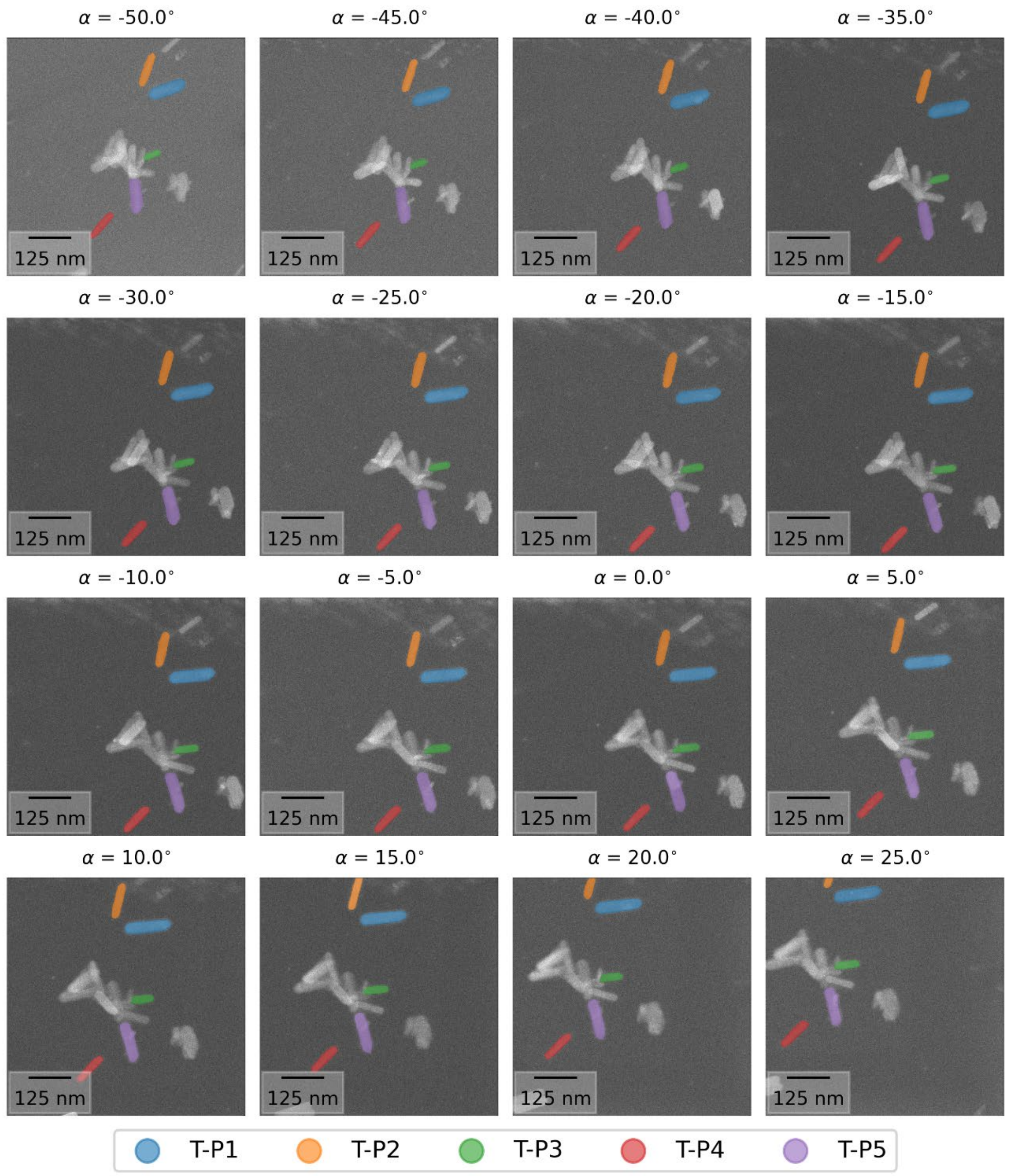

**Figure 3**. Particle tracking using the SAM2 algorithm on the DS-1 tomogram, plotted using representative images at 5$^{o}$ tilt intervals.

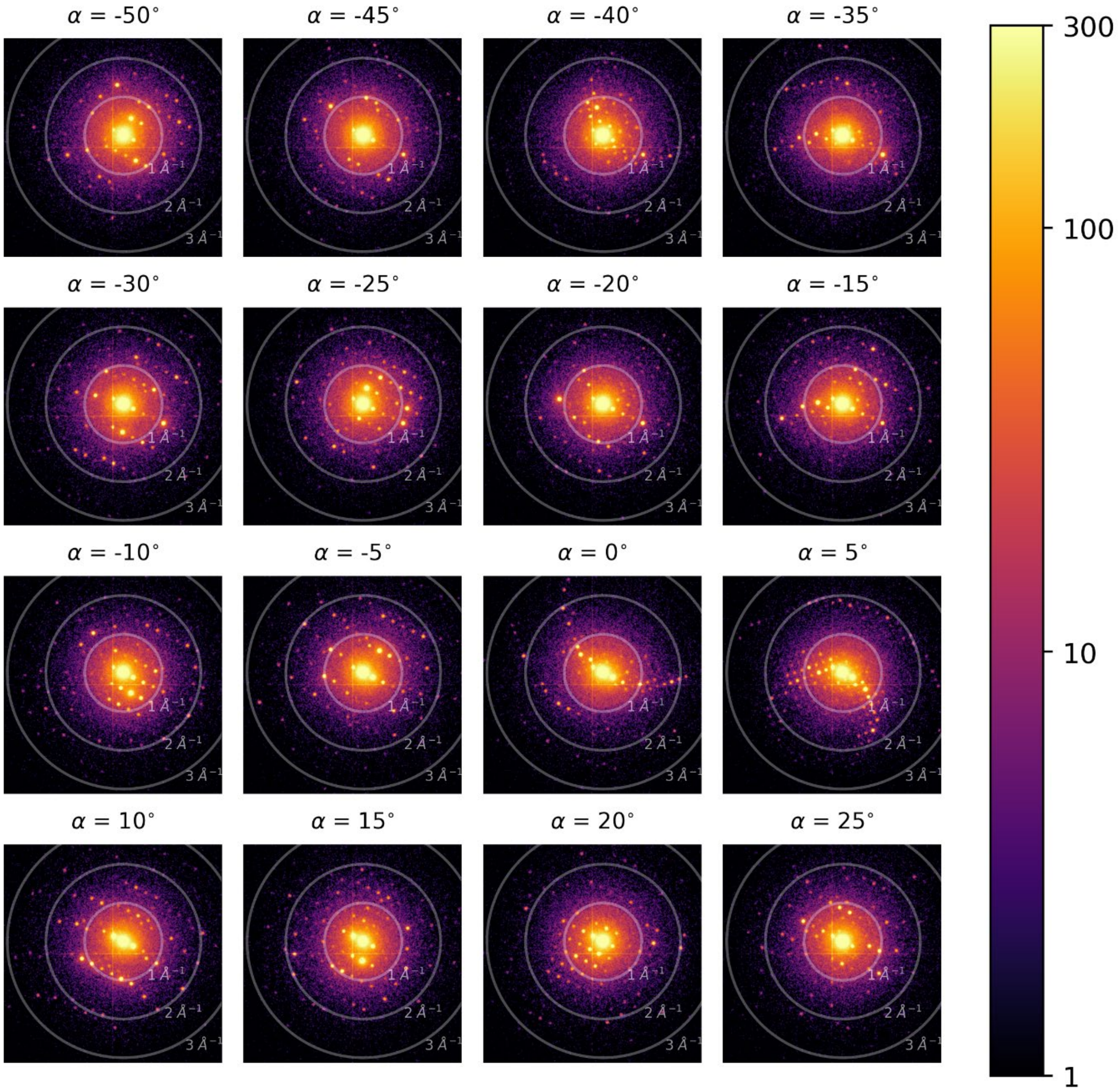

**Figure 4**. Representative 3D ED frames extracted from DS-1 corresponding to T-P1 in **Figure 3**. For better visualization, the intensities were plotted on a logarithmic scale and clipped at 300 counts.

Data reduction was performed successfully on all other particles with $R_{int}$ below 15% and $R_{meas}$ below 18% (**Figure 7**). The details of data reduction are listed in **Table S2**. Mainly due to dynamical scattering, $R_{int}$ and $R_{meas}$ are typically higher for 3D ED studies than single crystal X-ray diffraction, usually in range of 10 to 20 % and sometimes even reaching 25 to 30%.[32] The relatively low values found in the current study attest to the high integration quality of the datasets. The highest $R_{int}$ and $R_{meas}$ values were related to T-P1. This could be due to the larger size and different orientation of this nanorod increasing the probability of dynamic scattering. The symmetrically equivalent reflections in a dataset can originate from different frames, corresponding to a different path length through the crystal and different multiple scattering paths present in the frames. This results in intensity differences for reflections that are symmetrically equivalent.

The space group of the brookite phase is expected to be *Pbca* in the a > b > c configuration. This can be deduced from the reciprocal space sections calculated from the 3D ED datasets. On the T-P1 sections (**Figure 5** a, b, c), there were some violations of the *Pbca* reflection conditions, such as the visibility of $0kl{:}k \neq 2n$ and $h0l{:}l \neq 2n$ reflections. These violations are generally weaker than other reflections and occur mainly at lower-order reflections where the dynamical effects are expected to be higher. Such violations were generally reduced in the T-P3 sections (**Figure 5** d, e, f). This can be due to the different orientation of the particles and the slightly smaller particle size. This is opposed to X-ray diffraction techniques where typically larger particles are favorable.

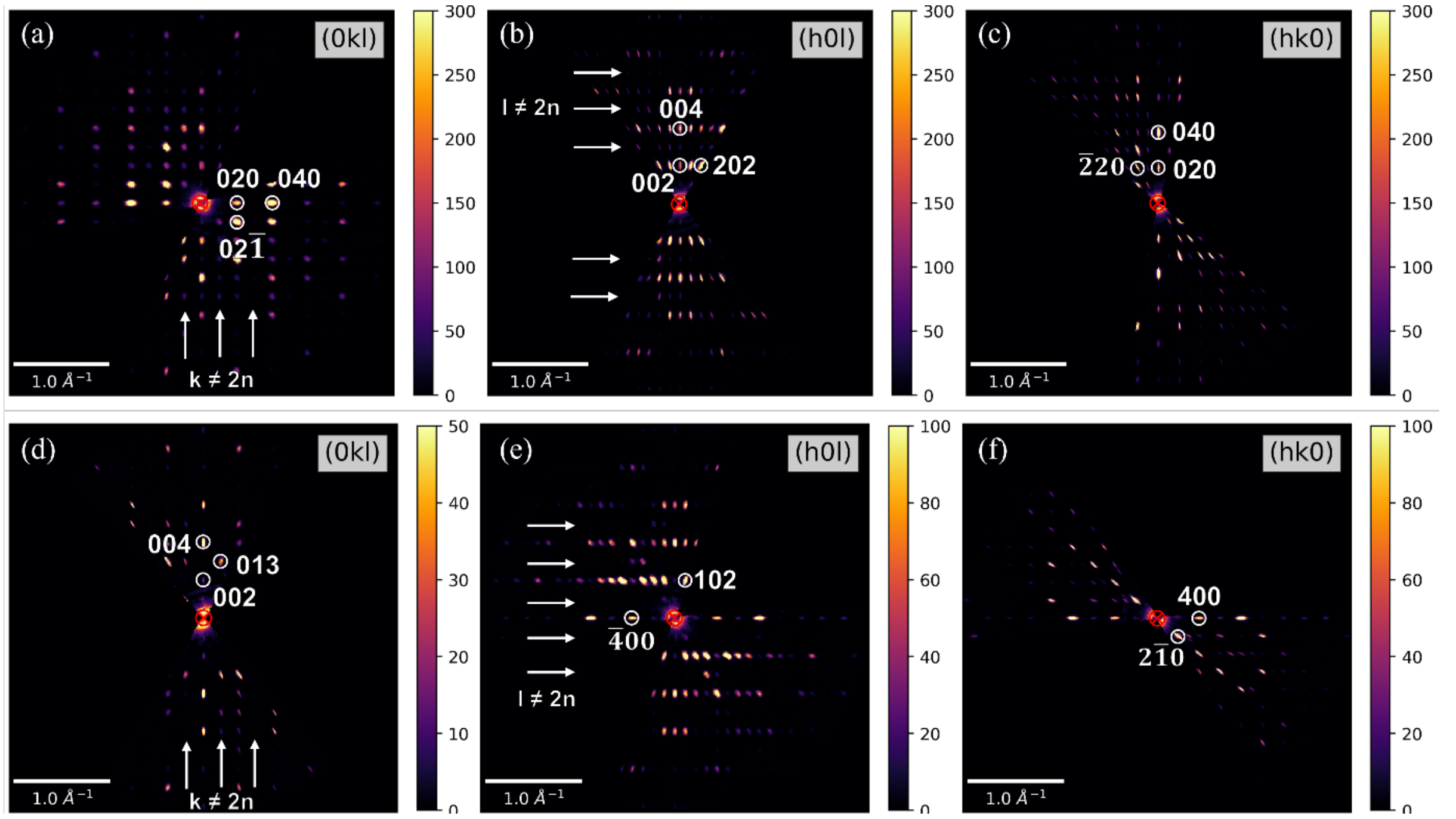

**Figure 5**. The main reciprocal space sections made from the 3D ED datasets of T-P1 (a, b, c) and T-P3 (d, e, f). The intensities are clipped at different counts for better visualization.

After integration, the violating reflections were discarded from the list of integrated reflections using Jana2020 software prior to the structure solution. Then, the structure was solved using the integrated intensities of the reflections by both SUPERFLIP and SHELXT.[33, 34] The solved crystal structure of the brookite sample is shown in **Figure S6**. Although both SUPERFLIP and SHELXT resulted in the correct space group in their solutions, SUPERFLIP usually resulted in extra oxygen atoms in positions unrealistically close to titanium. Such shortcomings are not unexpected in the initial solution from 3D ED, because, first, the completeness of the datasets was reported to be between 70 to 80 % by the PETS2 software. The remaining reflections are missing because of the missing wedge during the data collection. Second, the reflection intensities deviate from the kinematical approximation owing to multiple scattering. The structures solved using SHELXT were selected for the subsequent steps because they did not contain extra atoms.

In all cases, the refinements were straightforward, with all atoms freely refined, initially assuming isotropic ADPs, which is characterized by the $U_{iso}$ parameter representing the mean-square displacement of an atom from its equilibrium position. The details of the refinement results are listed in **Table S3**. For the kinematical refinements, all atoms in the refined structures were located within 0.03 Å of the reference brookite structure. This result is comparable to the best results previously obtained from conventional 3D ED on the same sample (see Table 1 in [3]). All atoms were refined to positive $U_{iso}$ values in the extracted datasets, except for T-P1 in the kinematical approximation (**Figure 6**). This means that the refinement resulted in meaningful physical values for the displacement of the atoms due to the thermal motion. T-P1 and T-P3 demonstrated the highest and lowest values of $R_{obs}^{kin}$, respectively (**Figure 7**). The negative $U_{iso}$ for O2 atom in T-P1 can be related to the high discrepancies in the intensities due to the dynamical effects. To evaluate the quality of the collected data with reference to the refined structures, the values of the three largest residual peaks in the difference electrostatic potential map (Δρ) above 3-sigma level were calculated and averaged based on the difference Fourier maps. Because the brookite structure was complete, these residual peaks simply indicate the remaining noise in the electrostatic potential maps owing to the kinematical approximation for the refinement.

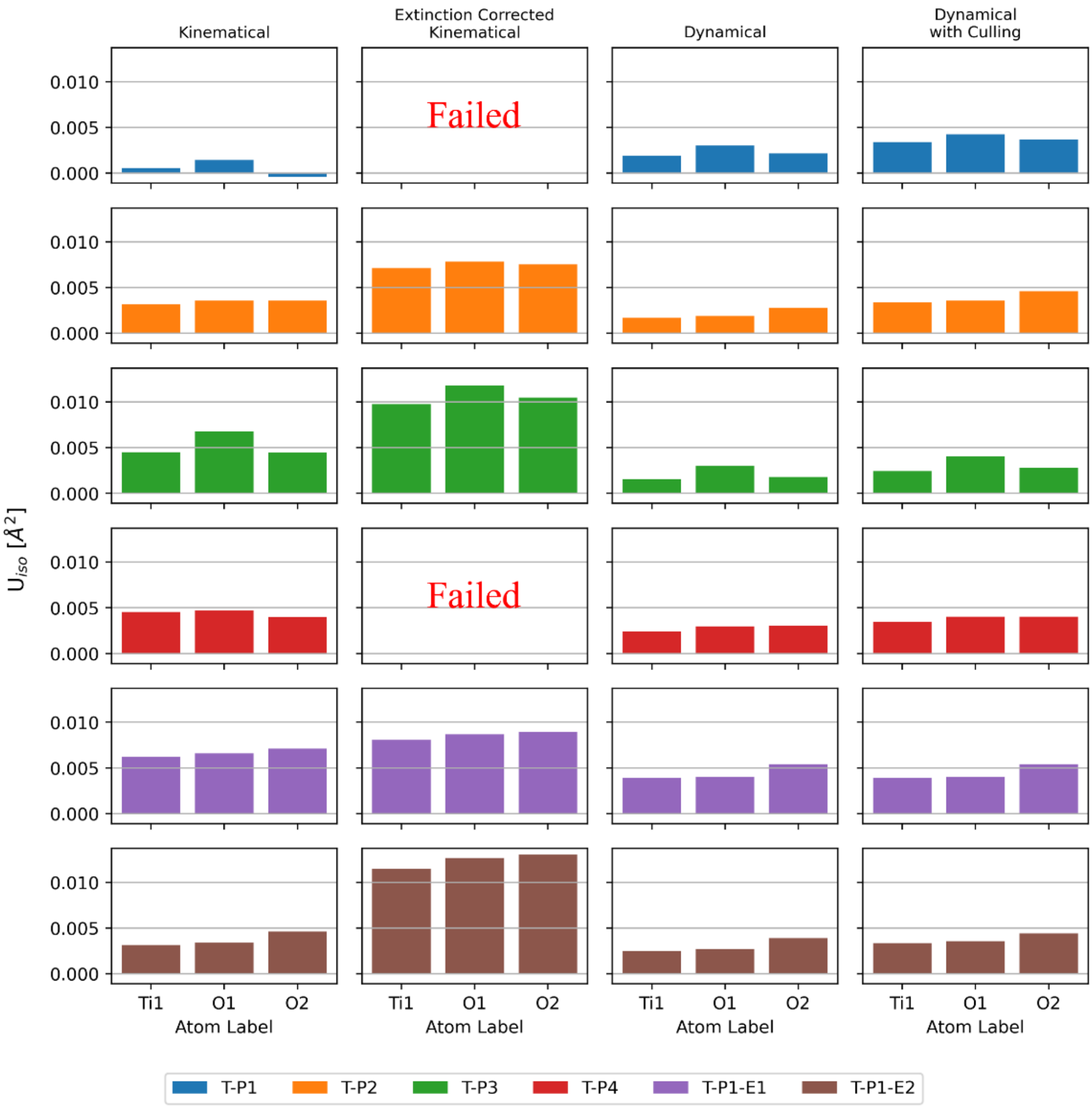

**Figure 6**. The $U_{iso}$ values of the atoms for each $TiO_2$ particle at different refinement stages.

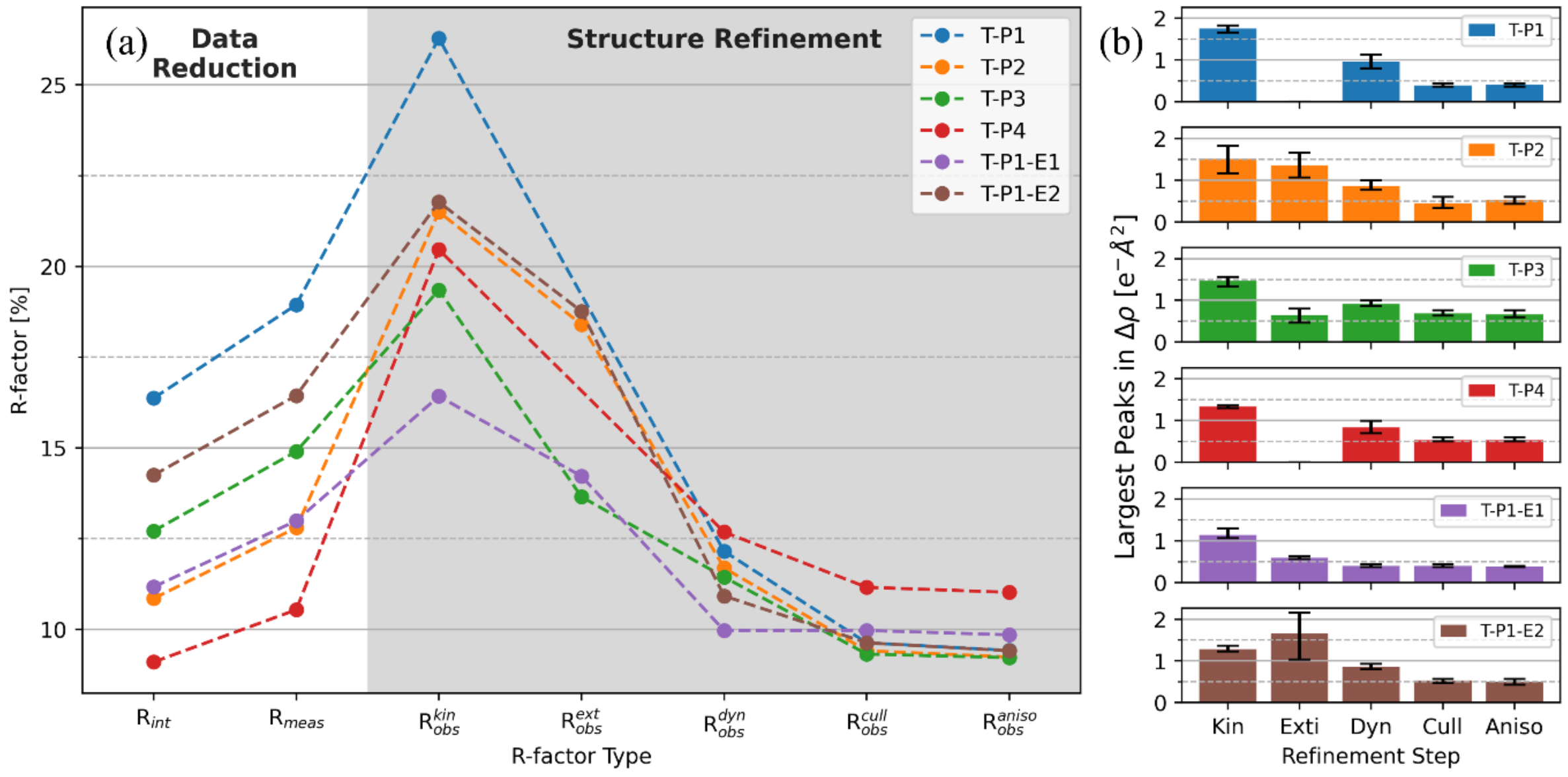

**Figure 7.** (a) R-factors during the data reduction and structure refinement stages; (b) mean values of the top 3 largest residual peaks in the difference electrostatic potential map at each refinement stage. The T-P1-E1 and T-P1-E2 curves correspond to the data obtained from only the edges, as explained further in this paper.

The addition of the extinction correction caused instability in the refinement process, sometimes leading to complete failure due to the divergence of the refinement. Nevertheless, this parameter significantly decreased the R-factors for some datasets. A closer look revealed that the $U_{iso}$ values were significantly and oddly higher in these datasets when extinction correction was introduced (**Figure 6**). This was also noticeable compared to the dynamical refinements performed later. Here, extinction correction was introduced using the SHELX implementation, which dampens the calculated structure factors. This formula aims to tackle primary and secondary extinction using an empirical approach. However, in electron diffraction, dynamical diffraction is a prevalent phenomenon even in thin particles, and it redistributes the intensities between strong and weak reflections.[35] Therefore, the improvements in the R-factors on application of the extinction correction occurs because of unphysical reasons, and a full dynamical regime better be applied.[35] This can also be observed in the changes in the largest residual peaks in the difference electrostatic potential map, which do not follow a consistent trend in the different particles studied.

Refinements that account for dynamical scattering effects, as for example implemented in Dyngo inside Jana2020 [36], address the redistribution of intensities between strong and weak reflections through the Bloch wave formalism.[8, 25] This relies on acquiring integrated intensities. The 4D STEM tomography approach presented here records 'static' ED patterns with partial integration as a result of the convergence of the beam. Additionally, it benefits from static frames acquired at fine tilt steps, a process that becomes equivalent to integration in the fine-slicing limit. This approach was first demonstrated using a fine-sliced static frame 3D ED dataset acquired with a 0.1° tilt step, for the dynamical refinement of the structure of cobalt aluminophosphate in [8]. While it has rarely been applied in practice for performing dynamical refinements, we exploit this method here in 4D STEM tomography.

As expected, applying dynamical refinement significantly reduced $R_{obs}$, on average from 21.9% ± 2.6% in kinematical refinement to 12.0% ± 0.5% in dynamical refinement. A decreasing trend was also observed for the $U_{iso}$ values of all atoms, except for T-P1, where the ADPs were slightly enlarged. Nevertheless, the largest residual peaks in the difference electrostatic potential map decreased drastically in all cases. After culling reflections that poorly matched due to non-crystallographic reasons, such as saturation of the detector, incomplete profile due to being on end frames or deviating rocking curves, the $R_{obs}$ decreased to 9.9% ± 0.7% on average, and the residual peaks decreased further. Interestingly, the $U_{iso}$ values were slightly larger for all particles than those obtained from the previous dynamical refinement step. In the final stage, anisotropic ADPs were introduced into the refinement. This did not have a noticeable effect on $R_{obs}$ or the residual difference Fourier peaks. All ADPs remained positively defined, and no unusually thin ellipsoids were observed. The refined structures were compared with the crystal structure obtained for the same brookite sample using synchrotron powder X-ray diffraction in [3] (**Table S5**). The values are in excellent agreement with the reference structure, with the differences in the positions all being below 0.01 Å after dynamical refinements, which is again comparable to the results previously obtained by 3D ED.[3] The bond valence sum (BVS) of the atoms also matched the values of the reference structure (**Table S4)**. These results reflect that excellent models with fine details were obtained from particles lying in close proximity on the TEM grid.

### *2.2.2. $CsPbBr_3$ Nanoparticles*

In this case, the CSRT tracker was used instead of SAM2 which is a faster and calculation-wise lighter tracker. **Figure 8** shows the particle tracking for $CsPbBr_3$ using the CSRT tracker in the DS-3 dataset, plotted at 5° tilt intervals. The tracker could still successfully follow the particles with a single starting point. No additional corrections were needed. **Figure 9** demonstrates the corresponding 3D ED frames extracted for particle C-P2 at those same specific tilt angles. Although a higher fluence was employed for this experiment, the diffraction resolution was 1.4 $Å^{-1}$. Compared to the brookite datasets, this lower resolution can be attributed to the small size of the particles and their lower crystallinity. A smaller condenser aperture was used for this dataset to reduce the convergence angle of the beam. This concentrates the scattered electrons in smaller reflections, facilitating their detection and integration, particularly for weaker reflections. Although the probe size increased in this case due to the diffraction limitation of the condenser aperture, the tracker successfully followed the small, blurred particles. Moreover, no noticeable resolution loss was observed over time for the extracted diffraction patterns. This shows that the experimental methodology successfully circumvents the resolution loss that would otherwise occur owing to contamination build-up.

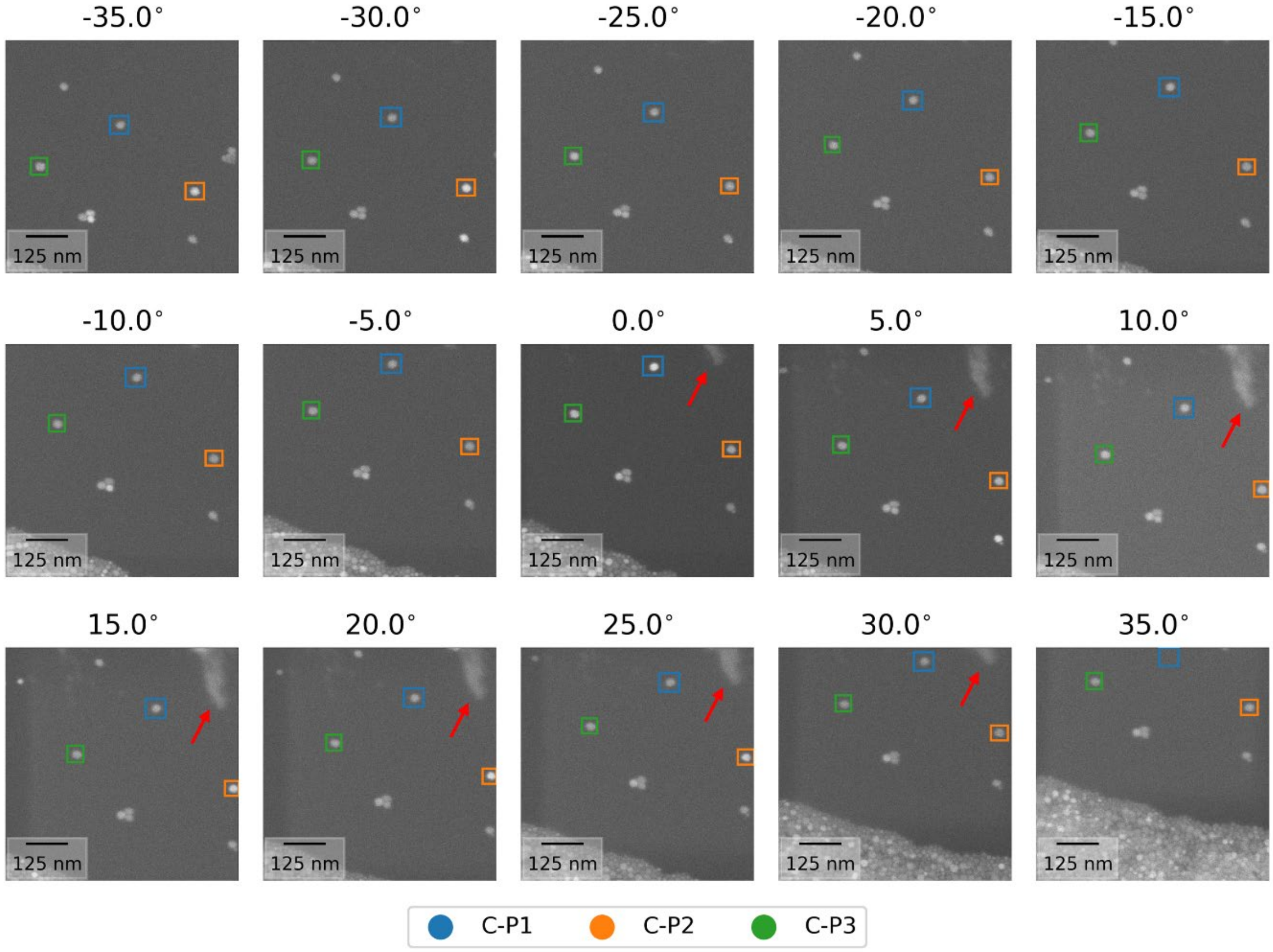

**Figure 8**. Particle tracking using the CSRT algorithm on dataset DS-3, plotted using representative images at 5° tilt intervals. The red arrows indicate the contamination growth at the beam-parking position.

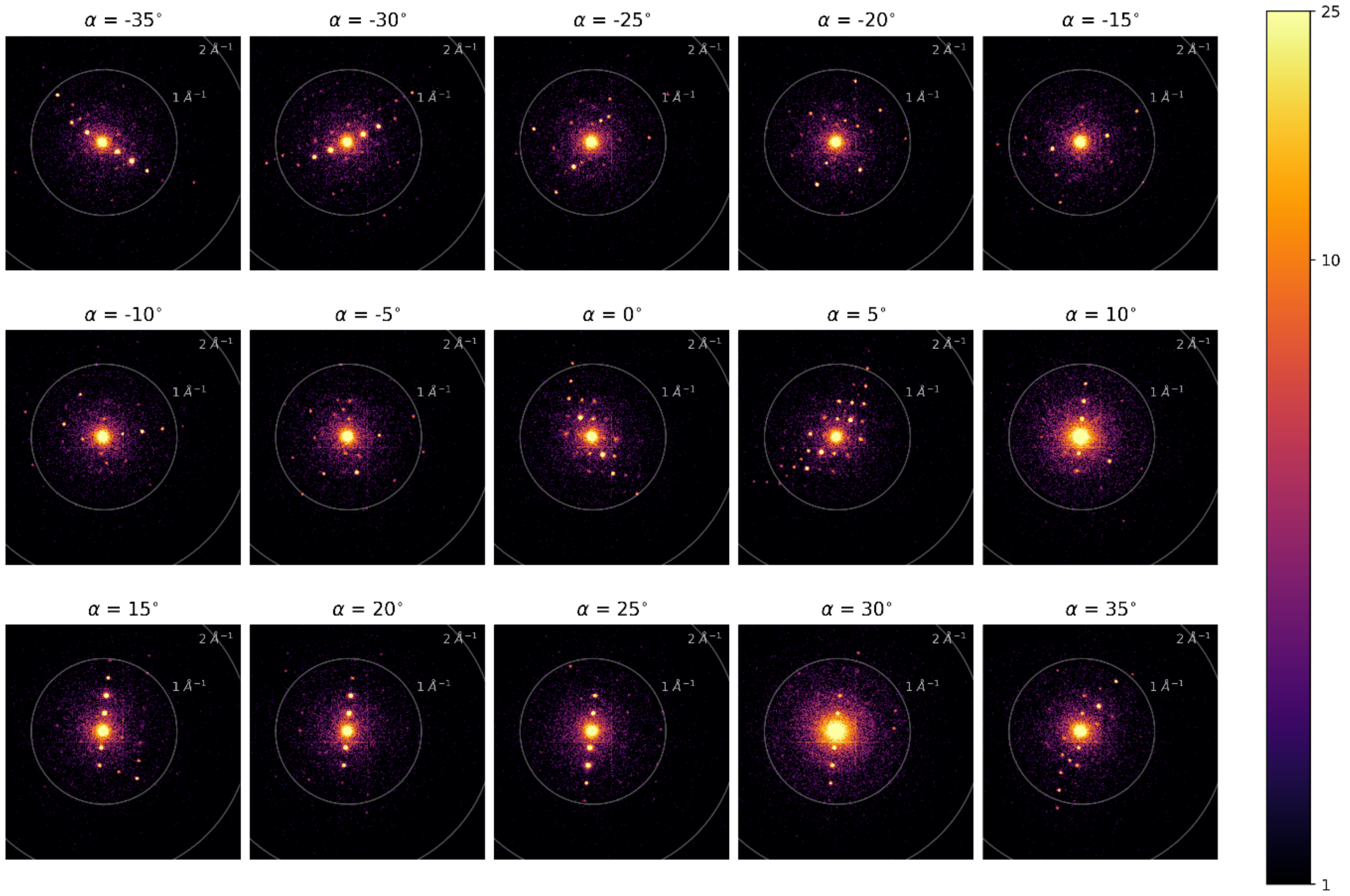

**Figure 9**. 3D ED frames extracted from 4D-STEM tomography for particle C-P2 at every 5° tilt. For better visualization, the intensities are on a logarithmic scale and clipped at 25 counts.

The expected unit cell of $CsPbBr_3$ was easily found using the extracted datasets with no sign of unit cell changes during the experiment. This indicates that the beam damage to this sample was effectively mitigated. $R_{int}$ and $R_{meas}$ were below 11% for all 3D ED datasets extracted from this tomogram (**Figure 10**). The detailed results are listed in **Table S4**. These results demonstrate the high quality of the datasets within their limited reciprocal resolution. The space group of $CsPbBr_3$ is expected to be *Pbnm* in c > b > a configuration (in accordance with ICSD ID #14608).[37] **Figure 11** depicts the main sections extracted from the C-P2 dataset. The reflection conditions could be derived from the reciprocal space sections and agreed with *Pbnm*, with only a few violations, which could be explained by multiple scattering. Both SUPERFLIP and SHELXT successfully solved the structure in the correct space group, although some atomic positions were either missing or incorrectly assigned to the wrong chemical species. The positions of missing atoms were identified in the difference Fourier maps following the initial kinematical refinement cycles and crystal chemistry considerations were able to correct the misassigned species. The atomic structure of $CsPbBr_3$ is shown in **Figure S7**.

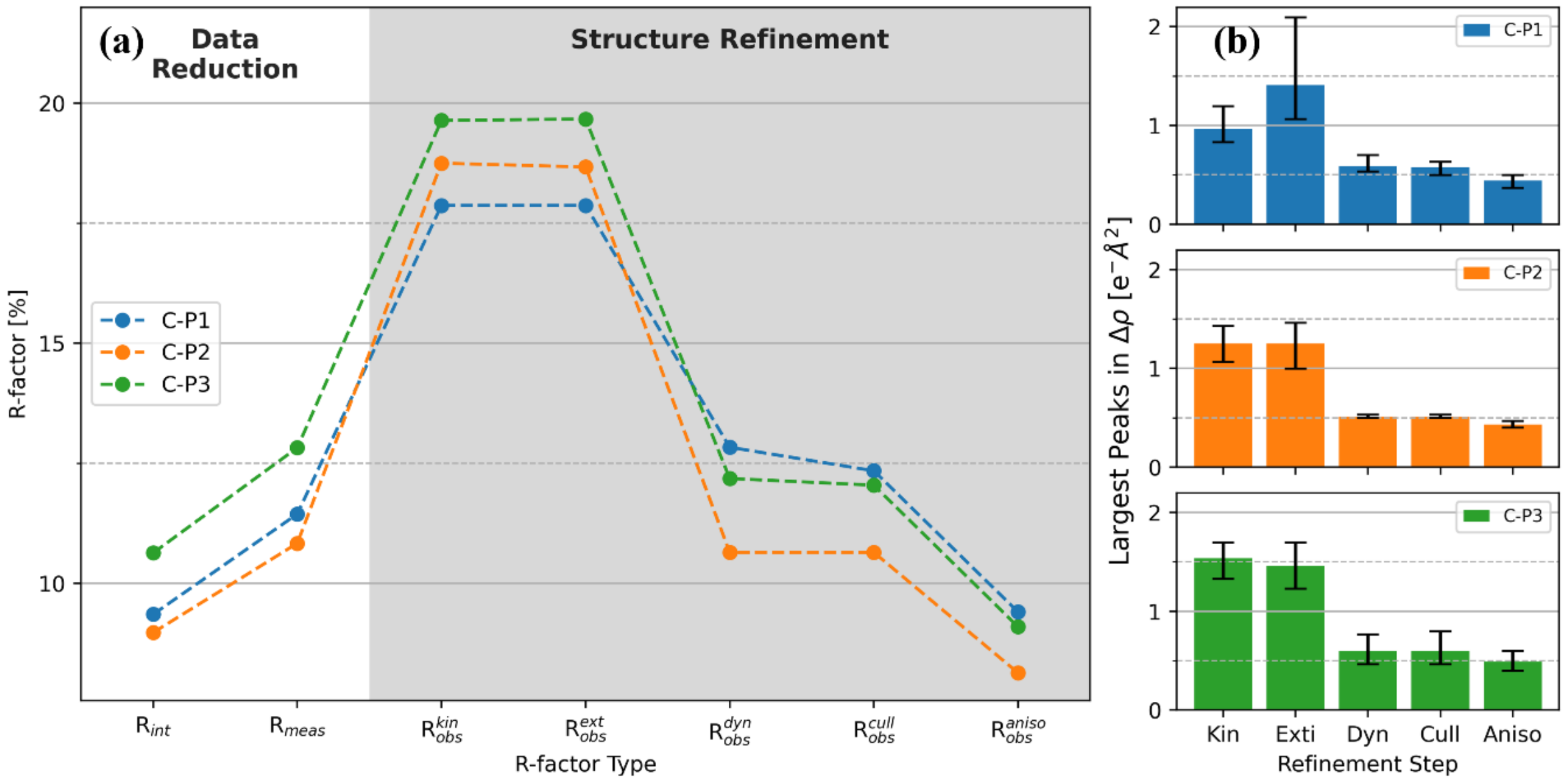

**Figure 10**. (a) R-factors during the data reduction and structure refinement stages; (b) mean of the top 3 largest peaks in the difference electrostatic potential map at each refinement stage for $CsPbBr_3$ particles.

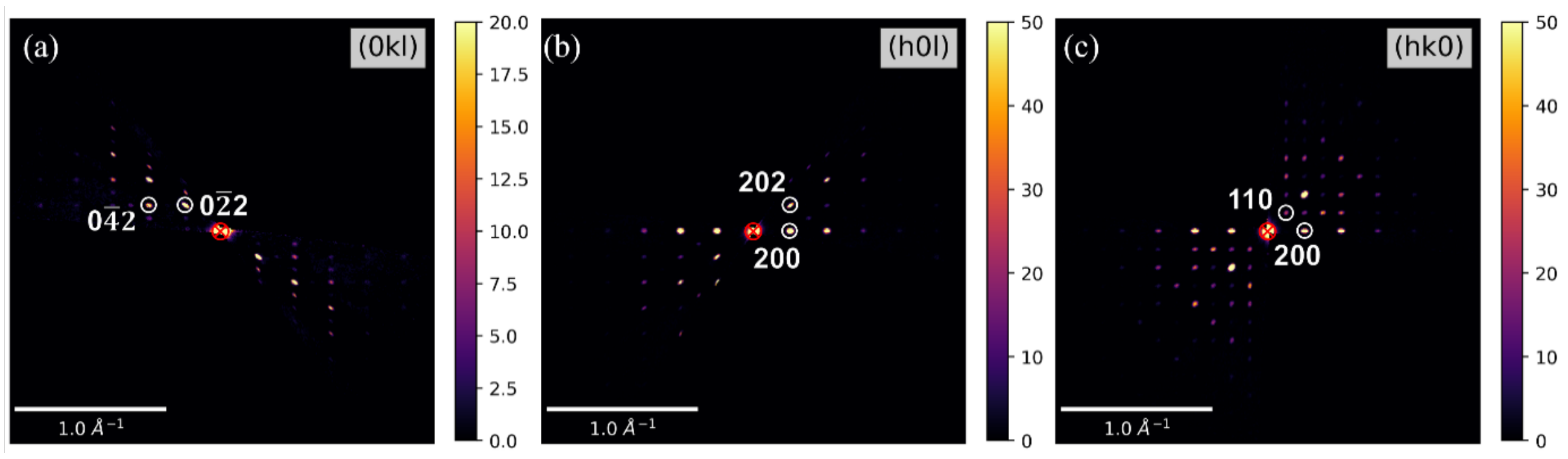

**Figure 11**. The main reciprocal space sections made from the 3D ED dataset of particle C-P2. The intensities are clipped at different counts for better visualization.

The refinement R-factors followed a similar trend as for the brookite particles (**Figure 10**). The $R_{obs}$ and the residual peaks in the difference electrostatic potential map were high in the kinematical refinement and later decreased significantly by dynamical refinement. The extinction correction was also unstable, similar to the results obtained in DS1. The detailed refinement results and calculated BVS are listed in Tables S5 and S6, respectively. Culling reflections excluded significantly fewer reflections than for the brookite particles. This is likely due to the better linearity of the detector in counting events for weaker reflections in $CsPbBr_3$.[24] In contrast to the brookite particles, the introduction of anisotropic atomic displacements noticeably decreased both $R_{obs}$ and the residual peaks. Thus, 4D-STEM tomography was successful in providing high-quality 3D ED datasets for the structure solution of the halide perovskites, overcoming difficulties such as beam damage, complications in locating small particles and contamination growth.

### *2.2.3. Sub-particle 3D ED*

Using 4D-STEM tomography, a 3D ED dataset can be extracted from any arbitrary region of a particle (**Figure 12**). For instance, edges can be of interest in the case of catalysts with possible bulk versus surface differences, for the study of core-shell structures, to follow transformations on the surface of nanoparticles during an ex-situ or in-situ TEM experiment, or simply to exploit the edge signal from thicker particles that are no longer electron transparent.[38] These possibilities represent significant advantages of the present approach over conventional 3D ED methods. The latter case was investigated for particle T-P1, which was the largest particle in the brookite dataset and exhibited high R-factors. To extract 3D ED frames from only the edges, the original segmentation masks were modified using an erosion algorithm. By subtracting the original masks from the eroded masks, only the edges remained. The thickness of the edge can be controlled by changing the kernel of the erosion (**Figure 12b** and **Figure 12c**). Finally, the diffraction patterns were extracted based on these edges from the corresponding 4D-STEM files.

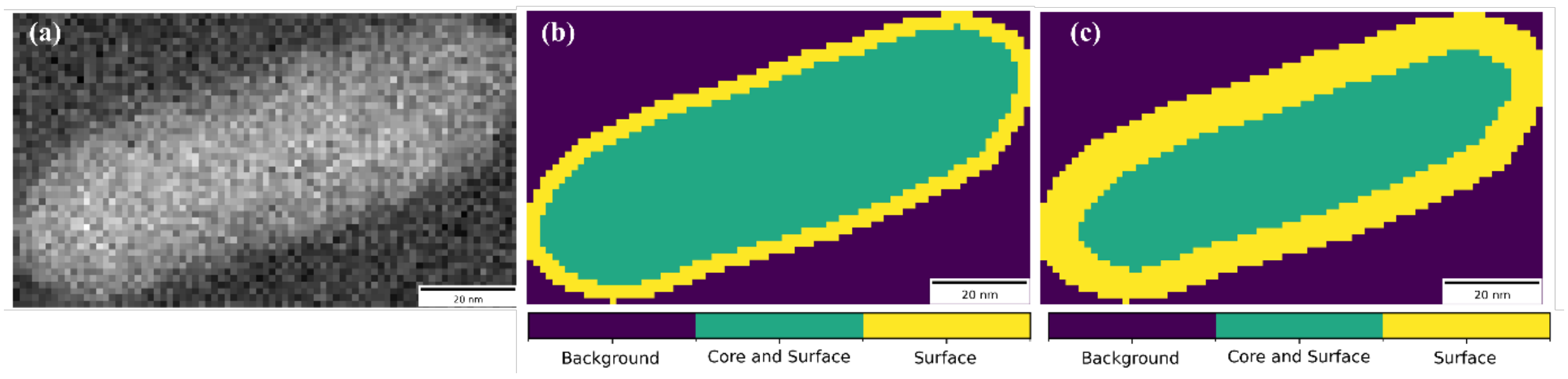

**Figure 12**. Custom masks and edge detection on T-P1 nanorod; (a) Reference nanorod, (b) thin edges resulting in dataset T-P1-E1, (c) thick edges - dataset T-P1-E2.

The R-factors for these datasets are shown in **Figure 7**. Compared to T-P1, the data from the whole particle, both $R_{int}$ and $R_{meas}$ decreased for these edge-only datasets, and T-P1-E1 reached values below 10%. This demonstrates lower discrepancies between the symmetrically equivalent reflections. Moreover, $R_{obs}^{kin}$ was remarkably lower for both new datasets. These better R-factors demonstrate a better match between the model and the observation, which we attribute to less multiple scattering at the thin edges, which is consistent with recent 3D ED results obtained for 10 nm indium tin oxide particles.[3] Correspondingly, the R-factors of the dynamical refinement were lower for both edge-detected datasets than for the dataset T-P1. This could be due to effects such as mosaicity, inelastic scattering, and thickness variations, which are decreased in such thin regions. Interestingly, T-P1-E1, which is thinner, has the lowest R-factors among all the Brookite datasets. Overall, the quality of the edge-detected 3D ED improves by lower dynamical scattering. On the other hand, it suffers from lower signal to noise ratio (**Figure S8**). Additionally, the edge might not be representative as it could be more strained or contain more defects. The balance between these two factors defines the success of the refinement process.

## 3. Discussion

### 3.1. To Be or Not to Be Converged

In the previous sections, we demonstrated that the combination of 4D-STEM tomography and object tracking is a powerful method for structure determination from challenging samples. This is mainly due to the nanometric spatial resolution, which provides direct observations over a wide FOV. In this method, the convergence angle of the beam determines the probe size, thus the spatial resolution. In the literature on 3D ED studies, a parallel beam is usually used because the aim of 3D ED is to collect the intensity of individual reflections; therefore, the opening angle of the beam is typically kept as low as possible to obtain sharp and separate reflections. However, targeting a parallel probe can result in larger probe sizes, considerably decreasing the spatial resolution in STEM. The probe size is influenced by several components, including the geometric size of the source ($d_g$), diffraction limit due to the condenser aperture ($d_d$), spherical aberration ($d_s$) and chromatic aberration ($d_c$). The detailed calculations are described in the supporting information and **Figure 13a** demonstrates the estimated probe size with respect to the share of each component in the final probe size. The diffraction limit plays the most significant role in peak broadening within the quasi-parallel regime. This results in probe sizes of a 2 – 3 nm with convergence angles between 0.6 – 1 mrad.

In addition to providing higher spatial resolutions, slightly larger convergence angles can also help to integrate the reflection intensities. In reciprocal space, reflections are observed in a diffraction experiment as the conceptual Ewald sphere cuts certain reflections at specific excitation errors. A convergent beam translates to an Ewald sphere having a thickness as large as the convergence angle.[39] Such thick Ewald spheres effectively integrate the reflection intensities within a range of excitation errors. Compared with a static tomogram with a parallel beam, this significantly improves the integration quality of higher-order reflections. This integrative quality somewhat mimics the integrative effect of the precession movement, with the difference that a hollow cone of wavevectors are integrated in precession, while a convergent beam covers the whole incident cone.

Nonetheless, there is a trade-off for such beneficial effects. As the convergence angle increases, the reflections enlarge and form disks, increasing the chance of reflection overlaps. This limits the size of the unit cells that can be studied with a certain convergence angle. **Figure 13b** demonstrates such limits at 200 kV with respect to the largest interplanar spacing (d-spacing) in the structure. The larger the convergence semi-angle, the smaller the unit cell should be to avoid disk overlaps. The details of the calculations, as well as the influence of the acceleration voltage, are described in the supporting information. Moreover, the intensity of the reflections is spread out under convergent conditions. The disks can even contain contrasts related to dynamical effects. Although the intensity spread helps in electron counting by direct electron detectors,[24] such patterns can produce errors in the peak detection process. However, such negative effects were not observed for the 3D ED datasets acquired on DS1 with a 1.2 mrad convergence semi-angle, translating to a 1.2 mrad or 0.14° integration in-between consecutive 3D ED frames. This effectively translates the data from this step-wise tomogram into a 3D ED dataset with an integration quality similar to that of continuous rotation or precession-assisted 3D ED data. To take this integration into consideration, the data was treated as a continuous rotation dataset in PETS2 with the convergence semi-angle as the tilt semi-angle.

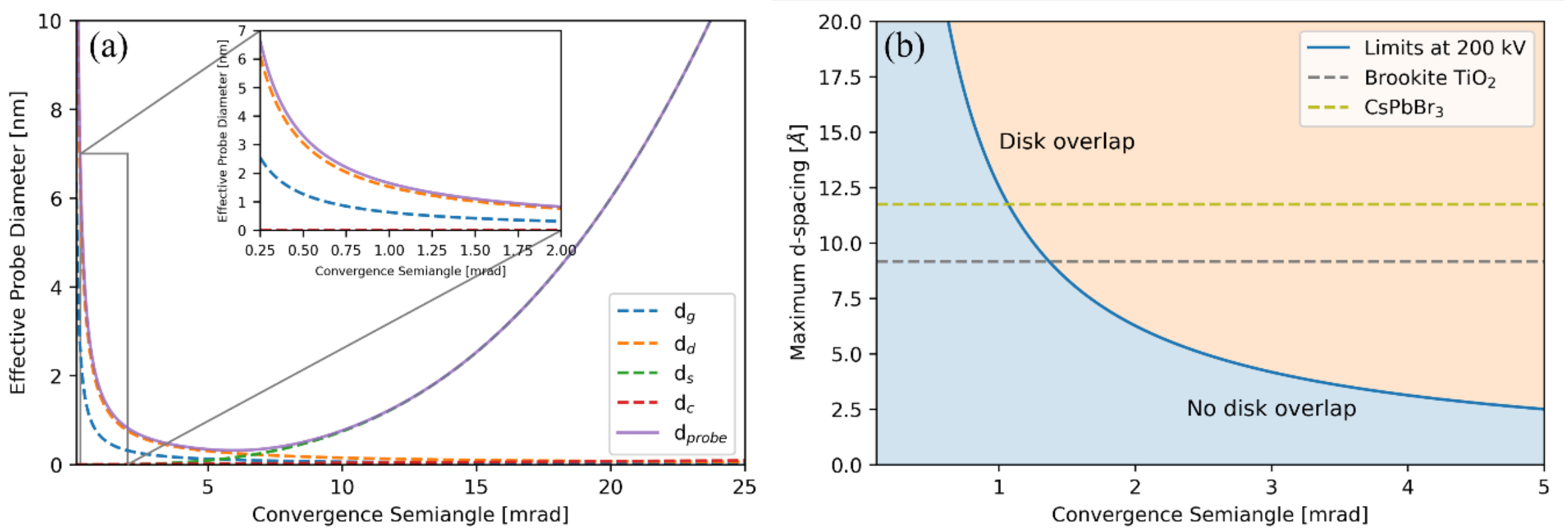

**Figure 13**. (a) The estimated probe size ($d_{probe}$) in STEM with respect to the aberrations in a non-corrected TEM at 200 keV. (b) the maximum d-spacing that can be studied without overlapping reflections.

## 3.2. The Duration of the Experiment

While a few hundred diffraction patterns are collected in conventional 3D ED methods in the TEM mode, this number can increase to several million frames in 4D-STEM tomography. Consequently, the duration of the experiment can be a concern. The total data acquisition time mainly depends on the scanning dwell time, real-space sampling, and angular sampling in the tomogram. **Figure 14** illustrates the estimated acquisition time for a range of these parameters. To avoid an experiment taking several hours, a trade-off should be found between the required structural information and these parameters.

According to **Figure 14**, the time of the experiment is usually expected to be longer than that of conventional 3D ED methods, which take just a few minutes. However, this does not increase the beam damage, as each individual scan point is only shortly illuminated during the entire experiment and the fluence can be adjusted to remain the same. The requirements for prior alignment steps are also relaxed in 4D-STEM tomography, decreasing the illumination time. For instance, the significance of a perfect eucentric height adjustment is relaxed because the FOV can be sufficiently wide (several micrometers) to include the particles of interest. Alternatively, the screening time can be minimized to a fast 4D-STEM scan to check the quality of the particles in the FOV and several 3D ED datasets can be obtained from one 4D-STEM tomogram.

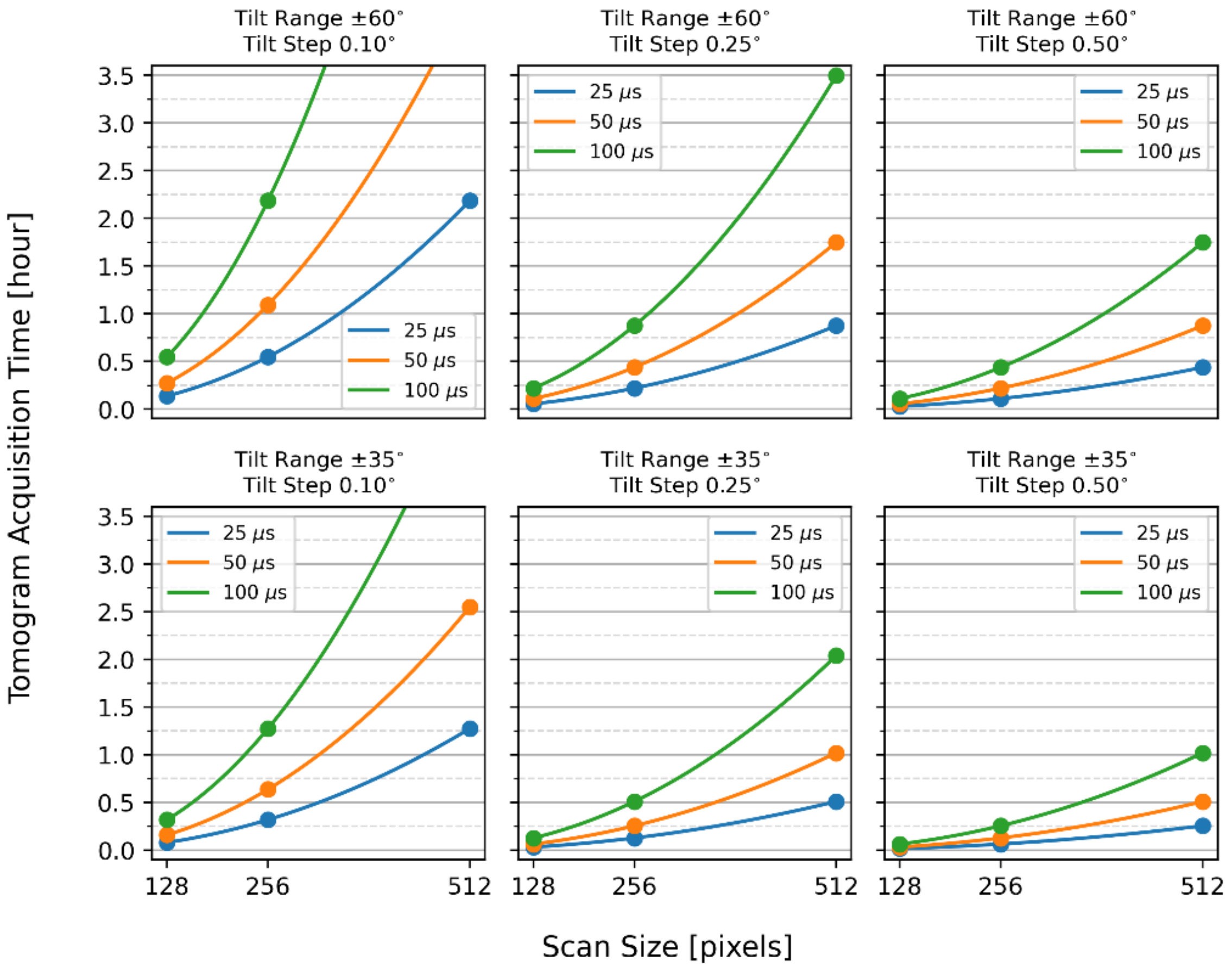

**Figure 14.** The duration of individual 4D-STEM acquisitions in a tomogram for a range of dwell times, scan sizes, tilt ranges, and tilt steps.

Among the mentioned parameters, the choice of scanning dwell time relies mainly on the scattering power of the sample, beam sensitivity, and detector limitations. These can be tested prior to the experiment using several 4D-STEM scans. The real-space sampling can be defined with respect to the probe size and FOV. The angular sampling can be more challenging to select. The tilt range is usually defined by the experimental conditions, such as the limitations of the TEM holder or shadowing of the particles by the edges of the TEM grid window. The tilt step defines how finely the diffraction space is sampled, determining the accuracy of the intensity profile of the reflections. A large tilt step causes large missing wedges in the collected intensity profile, leading to low partialities. Most importantly, the effectiveness, relevance, and justification for using dynamical refinements with static frames are only demonstrated when the reciprocal space is finely sliced. The more detailed structural information that is targeted, the more accurate sampling is thus required. A convergent beam integrates a portion of the intensity profile as wide as the convergence angle. Therefore, the more the convergence angle, the larger the tilt step can be. The crystal mosaicity and rocking curve width also play a role on how fine the sampling should be. The latter depends on the extinction distance which is directly proportional to the unit cell volume and inversely proportional to the structure factor of the reflection and the wavelength of the electron beam.[39] If the resolution is low, a higher tilt step can be used as it is easier to sample reflections of lower order. It is valuable to explore the optimal trade-off between fine tilt increments and reduced total data acquisition time. For instance, the results on the DS-3 dataset demonstrated that the desired sampling quality for a limited reciprocal space resolution could already be achieved using a 0.25° tilt step, corresponding to only 30% integration by convergence.

The effect of poor sampling by large tilt steps was investigated further on the brookite structure using the DS2 dataset. A 3D ED dataset was extracted from a single nanorod, and the frames in-between were gradually removed from the 3D ED dataset to evaluate the errors caused by sampling the integrated intensities at higher tilt steps. **Figure 15** displays the correlation between the integrated intensities of the initial dataset (0.1°) with under-sampled datasets having 0.2°, 0.5°, 1°, and 2° tilt steps. Although the intensities match well between the 0.1° and 0.2° tilt step datasets, they start to noticeably scatter from a linear correspondence for higher tilt steps, increasing the mean absolute error (MAE) by several orders. Additionally, when using higher tilt steps, several reflections are discarded from the list of reflections by the PETS2 software, denoted as unobserved or poorly integrated reflections. The number of discarded reflections amounted to 8%, 29%, and 74% of the initial reflections as the tilt step increased to 0.5°, 1° and 2°, respectively. It is worth noting that the reflections with relatively large d-spacing, i.e. low index reflections, also face severe errors upon enlarging the tilt step, and the errors are not confined to the reflections at larger distance to the center of the diffraction pattern. A 0.25° tilt step may represent the upper limit for acquiring 4D-STEM tomograms when the objective is accurate structure refinement that accounts for dynamical scattering effects. However, larger tilt steps can be employed for faster tomogram acquisition if the focus is solely on phase identification and structure solution. It must be noted that when frames are removed from the tomogram, although the SNR of each frame remains unchanged, the total fluence for the integration of a reflection is reduced. As a result, Figure 15 describes the combined effect of reduced tilt step and reduced Poisson statistics.

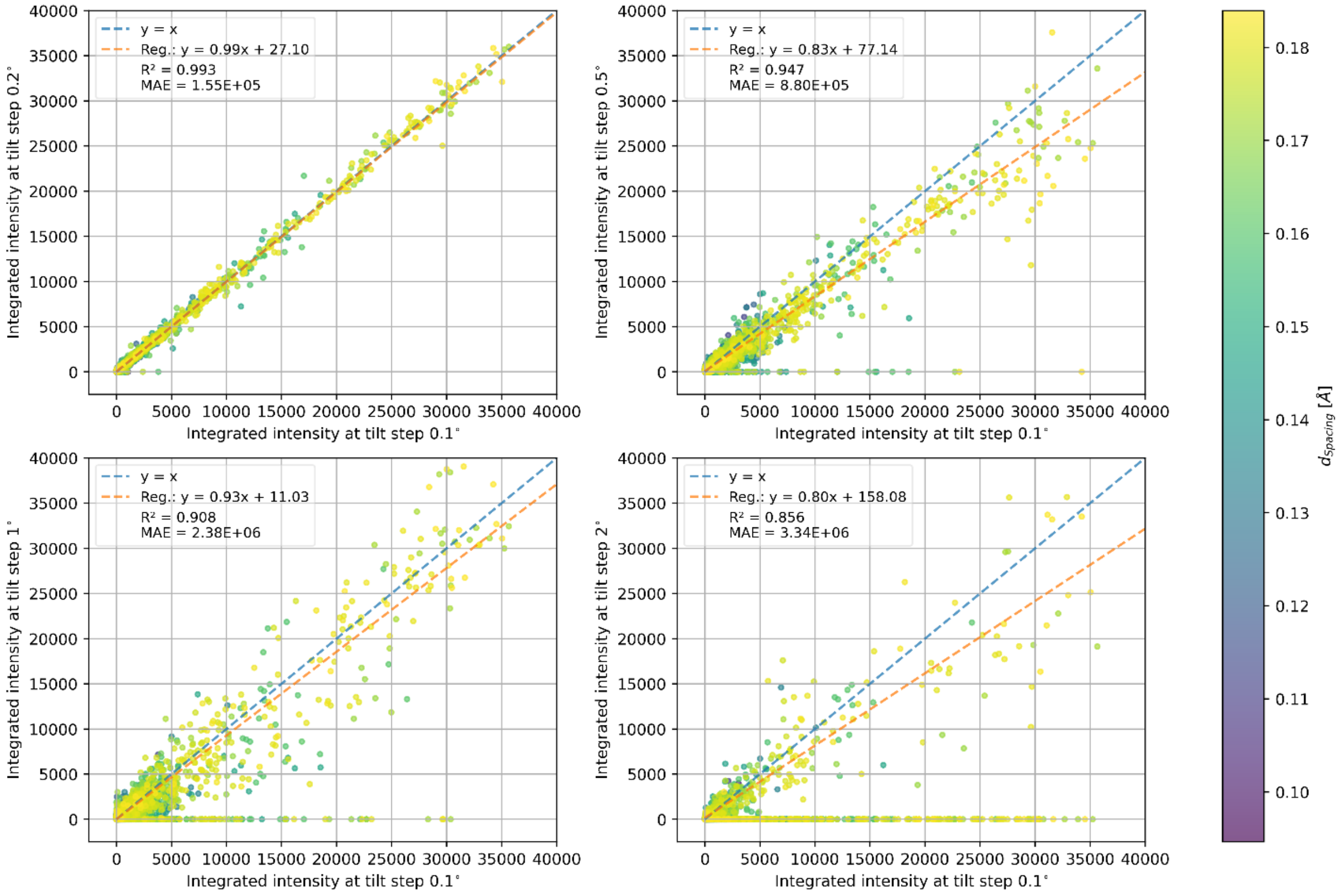

**Figure 15**. Correlation of the integrated intensities between datasets with different tilt steps made from the same frames. The reference dataset is the 0.1° tilt step dataset DS2, which is plotted on the x-axis. The regression was performed on all non-zero reflections at larger tilt steps. The points lying on the y = 0 line correspond to reflections that are not observed at higher tilt steps. The color of the dots represents the d-value of the reflection.

### 3.3. Data Management and Processing

We would like to emphasize the possibility of handling 4D-STEM tomography on typical modern desktop computers. This certainly allows for the adoption of the technique by a broader audience. Such a goal can only be achieved with proper management of the processing steps. After data collection, a preliminary step in data analysis is to establish a protocol to efficiently handle files. To provide an idea of the dimensions of such a challenge, the tomograms used in this study consisted of 400, 1000, and 280 4D-STEM files for the DS-1, DS-2, and DS-3 datasets, respectively. For instance, in the case of DS-1, the size of each raw event file amounted to 1.4 GB on average, reaching over 500 GB for the entire tomogram. A similar dataset can be multiple times heavier in the case of using frame-based detectors, as the whole frame is written on the hard drive at a specific bit depth, in contrast to the event-based detector used here recording a list of individual events. Thus, data management plays an important role in efficient calculations in data analysis and storage.

In the proposed workflow, the calculation of the virtual image and the extraction of diffraction patterns from the ROIs are computationally heavy. The rate of the operations can depend on different factors, such as the data format, scan size, beam current, or number of pixels for ROI calculations. Therefore, it is difficult to draw a solid conclusion regarding the ideal data format. A more detailed discussion is provided in [24]. In this study, as the size of the tomograms was typically large, addressing the data storage issue was more important than the subsequent calculations. Therefore, working on compressed data formats was preferred, even though they might not be ideal for the calculations. Such operation typically decreases the size of the files by 70% to 80%, which makes handling the files much easier. In this study, a compressed HDF5 format was selected for storage. The rate of conversion, compression, and writing of the file depends on the compression factor, number of CPU processors, available memory for the computations, and type of hard drive. Employing a multi-processing approach on a desktop computer equipped with an Intel Core i7-10700 CPU with 16 logical processors having 15 GB free memory, the whole procedure took 2 to 4 h for the datasets in this study. A more appropriate approach would be to perform the conversion and compression on-the-fly during data acquisition at the microscope.

The first step in the data analysis was to calculate the navigation images. We performed this step simultaneously during file conversion to save calculation time. For the extraction of frames from the tracked and segmented regions, loading the bulky 4D-STEM files might not be possible and is inefficient. Thus, the analysis steps were performed using Dask arrays which load small chunks of data in the memory for executing calculations.[40] In this study, the extraction of the frames took approximately 25 s per 4D-STEM file with a 512×512 scan size. Using up to seven processors, each 3D ED dataset of the brookite sample was extracted within 25 min from DS-1. This process can be further enhanced by optimizing the chunk size in HDF5 files or migrating to a data format for faster Dask operations, such as Zarr.[41]

Another significant data processing step is object tracking. Although both trackers performed well on the presented datasets, they exhibited some differences. In SAM2, two points were used to specify the foreground and background of the object, and this model tracked the object, resulting in segmented regions per frame. In the case of CSRT, similar to other trackers in OpenCV, a rectangular region was used as input, and later a segmentation algorithm, such as the Otsu threshold [42] or Li threshold [43] was used to detect the foreground. The SAM2 tracker was generally much more robust in tracking particles when there were sudden jumps in the movement of the object during tomography experiments. It also provides a straightforward way to select an entire particle in a agglomerate. The CSRT tracker might require more sophisticated segmentation steps in this condition to isolate the particle from the surrounding particles. A good example can be the agglomerate on DS-3 where three particles are attached together. While tracking these particles can be a challenge with CSRT, SAM2 can easily follow these along the tomogram (**Figure S9**). Nevertheless, CSRT is more straightforward for tracking a part of a large particle because it confines the region during tracking. With respect to the computation cost, the SAM2 tracker is very expensive and usually requires GPU acceleration, while the CSRT tracker is very fast and light. Overall, it should be noted that the proposed workflow can be adjusted for a specific sample, for instance, by applying a different tracker or segmentation procedure or by employing a 2-step process, combining tracking algorithms to have a blend of the advantages of each algorithm.

### 3.4. Benchmarking Against State-of-the-Art

As demonstrated, the combination of automated 4D-STEM tomography, fine-slicing the reciprocal space, and object-tracking during post-processing circumvents many challenges for obtaining 3D ED datasets on complicated cases such as particle clusters and agglomerations. The challenges of particle tracking at the microscope are completely eliminated, and there is no need for predicting the stage behavior or particle position during the data collection. Diverse algorithms can be applied during post-processing to reach the desired 3D ED datasets. This only requires the desired particle to stay in the FOV which can be checked at the beginning of the experiment using HAADF-STEM images. The FOV can be easily several micrometers wide.

The advantages of 4D-STEM tomography do not confine to higher success rates on agglomerated samples and simpler particle tracking. Thanks to the digital formation of the diffraction frames and the applied segmentation, the diffraction signal has minimal interference from the surrounding TEM grid support known as background noise. This is a significant improvement in signal strength particularly for small, beam-sensitive nanoparticles where the fluence must be decreased which already reduces the signal strength inevitably and undesirably. The $CsPbBr_3$ sample presented in this work can easily form Pb clusters due to beam damage [44], thus it obliged employing minimal fluence. This would complicate the observation and 3D ED experiment setup for such small particles in the TEM without damaging them. Nevertheless, it was possible to acquire a full dataset without any sign of structural damage on several particles. It should be noted that it is also possible to adopt a scan size with smaller scan points, resulting in an under-sampled FOV in 4D-STEM tomography. This can effectively reduce the fluence on the particles, with the price of a reduction in the signal strength and somewhat pixelated images.[17] It can also be claimed that the need for crystal growth is even further relaxed in this approach as the technical difficulties for data acquisition, such as particle tracking, is overcome and stronger signals can be collected from smaller particles.

The nanometric, spatial resolution of the 4D-STEM datasets provides valuable real-space information, and even the opportunity for real space 3D reconstructions.[18] Each 4D-STEM tomogram can provide 3D ED datasets as well as 3D view on many particles in the FOV. Compared to conventional tracking approaches in 3D ED, no information is lost by unpredicted circumstances and the rich 5-dimensional data can provide structural and visual information on any particle and even custom sub-particle regions, such as edge-detected 3D ED. Notably, even if some features or particles were unnoticed during the microscopy session, they can be picked up during the data analysis, thanks to flexible virtual images, and examined structurally.

In order not to sacrifice the quality of the 3D ED datasets, a fine-slicing regime can be adopted with a slight convergence angle ensuring detailed sampling of the reciprocal space. This removes the necessity for utilizing precession, effectively simplifying both the microscope requirements and the experiment operation, as well as reducing the additional hardware costs of the microscope. The fine-slicing approach can raise concerns on the size of the datasets; however, we demonstrated that by utilizing a Timepix3 detector, the data can easily be handled by a modern desktop computer. The datasets can be compressed further and analyzed by open-source distributed computing techniques which eliminates the need for any high-performance computing system. In a separate, subsequent publication, we will provide software with a user interface to eliminate the need for any programming skills for analyzing such datasets.

## 4. Outlook

To unlock the full potential of 4D-STEM tomography, we envision several improvements. A main drawback of the method in its current form is that the total time of the experiment is longer than that of conventional 3D ED. A possible solution is to increase the scanning speed and tackle the consequent reduction in signal strength using other approaches. The simplest solution is to increase the current. This requires detectors able to work at higher fluences without sacrificing speed, such as Timepix4 technology.[45] Furthermore, it is possible to reduce the acceleration voltage of the TEM to increase the scattering cross section of the electron beam with matter.[46] Therefore, the signal is expected to be several times stronger at, for instance, 30 kV compared to 200 kV, although the differences in beam damage mechanisms at lower acceleration voltages and thickness limitations should be considered. Regardless of the scanning dwell time, a smart scanning strategy can be highly efficient for reducing the number of scan points and thus speeding up the acquisition.[38, 47, 48] By scanning only the particles, and not the empty surrounding area, the time of the experiment decreases remarkably. For instance, on the datasets presented in this study, such an approach can decrease the number of scanning points by 90%. Such a remarkable reduction would close the gap in acquisition time per dataset between 4D-STEM tomography and CRED. Additionally, such speed-up allows acquisition of 4D-STEM tomograms over a wide FOV providing statistical results over an unknown sample. The overlap of the domains or particles is another major issue in all techniques relying on single-crystal analysis, and is not fully addressed here either, because areas with overlapping patterns are avoided instead of analyzed. In our proposed workflow for 4D-STEM tomography, the segmentation approach functions only based on the contrast of a region. Combining nanocrystal segmentation based on diffraction patterns [49-52] with 4D-STEM tomography would lead to an individual 3D ED dataset for each existing phase in the FOV if the individual lattices can be detected. However, in the case of overlapping reflections, the currently implemented approaches do not yet untangle the reflection intensities into the correct intensities per lattice. Untangling all crystal structures within multiphase particles with the correct partial intensities and a clear record of their location in the particle would be of high interest for following phase transitions.

## 5. Conclusion

We have demonstrated that combining 4D-STEM tomography with object tracking provides a convenient and effective way to solve and refine the crystal structure of nanoparticles under conditions where conventional 3D ED faces significant challenges. This method overcomes the limitations related to particle size, agglomeration, and beam sensitivity, and enables the extraction of regional structural information from multiphase or multidomain particles. This extends the applicability of 3D ED to previously inaccessible systems. By automating the data acquisition procedure, we managed to acquire hundreds of tomograms with small tilt steps and reach a high integration quality of the reflection intensities in 3D ED. Utilizing the effect of integration by slight convergence angles, this method enabled not only the structure solution, but also detailed structural information, such as refining the anisotropic ADPs of the atoms, by implementing dynamical refinement. Because many parameters are involved in a 4D-STEM experiment, a roadmap was drawn to demonstrate how to design the experiment and select each parameter to answer the structural question at hand. All these were provided by a workflow that allows processing of these heavy 4D-STEM tomograms with a conventional desktop computer. This facilitates the use and adoption of the technique by more users without the need to access high-performance computing facilities.

## 6. Experimental

### 6.1. Materials

Two samples were used in this study to demonstrate the applications of the proposed technique: brookite $TiO_2$ and $CsPbBr_3$. Brookite nanorods were used as reference samples to allow comparison with conventional 3D ED. 3D ED and synchrotron studies on this sample, as well as its synthesis, have been published in [3]. The brookite particles were nanorods with a range of sizes, all below 130 nm in length and 35 nm in width. Individual brookite particles are labeled as "T-Pn" standing for "$TiO_2$ – Particle n". The sample was dispersed in ethanol using bath sonication for 5 min before drop-casting onto TEM copper grids with an ultrathin, continuous carbon support (Electron Microscopy Sciences).

$CsPbBr_3$ nanoparticles are labeled as "C-Pn" standing for "$CsPbBr_3$ – Particle n". $CsPbBr_3$ nanoparticles were synthesized using a two-step process involving Cs-oleate preparation followed by hot-injection synthesis. The Cs-oleate preparation protocol was adapted from [53]. $Cs_2CO_3$ (390.8 mg) was loaded into a 100 mL three-neck flask along with 1-octadecene (18 mL, ODE, 90%, Aldrich) and oleic acid (2 mL, OA, 90%, Aldrich). The reaction mixture was degassed for approximately 1 h at 120 °C. The solution was then collected under hot conditions in a nitrogen-filled vial. Before injection, the vial was heated to 100 °C to dissolve the Cs-oleate in ODE. The second step was followed with slight changes in the parameters compared to the protocol from [53]. PbO (44.6 mg, 99.999%, Aldrich), phenacyl bromide (119.4 mg, 98%, Aldrich), OA (1 mL), and ODE (5 mL) were loaded into a 25 mL three-neck flask and degassed. The atmosphere was switched to nitrogen, the temperature was increased to 220 °C, and oleylamine (0.6 mL, OLA, 80-90%, Acros Organics) was injected. The solution became red within a few minutes and then gradually turned yellow after approximately 20 min. Subsequently, Cs-oleate (0.5 mL) was injected into the yellow solution and annealed for 15 min. The sample was collected with ice-quenching. The solution was centrifuged at 1200 rcf for 5 min, and the precipitate was re-dispersed in 2 ml of toluene. The solution was centrifuged a second time at 1200 rcf for 3 min, and the precipitate was discarded. The resulting suspension was severely diluted and drop-cast on the TEM grids before the experiments.

### 6.2. 4D-STEM Experiments

The experiments were performed using a Tecnai G2 TEM microscope (Thermo Fisher Scientific) operated at 200 kV using a Fischione tomography holder. To test different probe configurations, a custom-made C2 aperture was used which had a range of holes, including 10 and 20 μm diameter holes. The C1 aperture was used to select the desired hole in the C2 aperture and cover the rest of the holes. This microscope was equipped with a Timepix3 based CheeTah T3 Quad Detector (Amsterdam Scientific Instruments) with 512 × 512 pixels. A Quantum Detectors scan engine was used to perform the scans and send the required triggers at the beginning and end of each scanned line to the detector.

The 4D-STEM scans were performed using *evenTem* software suite which controls both the detector and scan engine.[24] To automate the tomography experiment, the microscope was controlled via the *temscript* package, a Python wrapper for the scripting interface of Thermo Fisher microscopes.[54] This functionality was integrated into the *evenTem* suite as an additional widget.

### 6.3. 3D ED Data Reduction and Structure Solution

The data reduction of the 3D ED datasets was performed in PETS2 software [6] in two stages. The first stage included peak finding, tilt axis refinement, unit cell determination, and the integration of the intensities. The second stage included distortion refinement [55], optimization of the frame geometry, and final integration of the intensities. For brookite samples, distortion refinement was performed using fixed cell parameters obtained from the synchrotron powder X-ray diffraction results in a previous study on the same sample.[3] For the $CsPbBr_3$ sample, distortion refinement was performed by restricting the data to the orthorhombic crystal system. To include the integration by the convergence angle, a continuous geometry was selected, setting the convergence semi-angle as the tilt semi-angle on PETS2. Structure solution was attempted using both SHELXT [34] and SUPERFLIP [33], with the better initial model being selected for subsequent refinements. Both software programs were employed from inside Jana2020 software suite.[36] The subsequent kinematical and dynamical refinements were also performed using Jana2020. The observed structure factors ($F_{obs}$) were used for least-squares optimization during the refinement. The detailed parameters are listed in the Supporting Information, and the final structures are provided in CIF format as supplementary files. The remaining peaks in the difference Fourier maps (reported as charge in Jana2020) are scaled by a factor of 3.32 to achieve values in $e^{-}Å^{2}$.

## 7. Acknowledgement

S.G. and J.H. would like to acknowledge the fruitful discussions with Alexander Zintler, Morteza Behrooz, Matthias Quintelier, Nadine J. Schrenker, Sepideh Rahimi, and Petr Brazda. We would like to also acknowledge funding from Research Foundation Flanders (FWO, Belgium) project SBO S000121N (AutomatED), SBO 1SHA024N and G069925N; the Research Fund of the University of Antwerp through projects BOF TOP 38689 and BOF/SEP 53221. This study is also funded by the European Union's Horizon 2020 research and innovation programme under the Marie-Sklodowska-Curie grant agreement No 956099 (NanED – Electron Nanocrystallography – H2020-MSCA-ITN). A.A and J.V acknowledge funding from the European Union's Horizon Europe programme under grant agreement no. 101094299 (Impress) Funded by the European Union. Views and opinions expressed are, however, those of the authors only and do not necessarily reflect those of the European Union or the European Research Council Executive Agency. Neither the European Union nor the granting authority can be held responsible for them. This work is supported by ERC grant (REACT, 101199099).

## 8. References

1. Gruene, T. and E. Mugnaioli, 3D Electron Diffraction for Chemical Analysis: Instrumentation Developments and Innovative Applications. Chemical Reviews, 2021. **121**(19): p. 11823-11834.

2. Gemmi, M., et al., 3D electron diffraction: the nanocrystallography revolution. ACS Central Science, 2019. **5**(8): p. 1315-1329.

3. Cordero Oyonarte, E., et al., 3D Electron Diffraction on Nanoparticles: Minimal Size and Associated Dynamical Effects. ACS Nano, 2025: p. 20599–20612.

4. Plana-Ruiz, S., et al., Fast-ADT: A fast and automated electron diffraction tomography setup for structure determination and refinement. Ultramicroscopy, 2020. **211**: p. 112951.

5. Kolb, U., Gorelik, T., Kübel, C., Otten, M.T., Hubert, D., Towards automated diffraction

tomography: Part I—Data acquisition. Ultramicroscopy, 2007. **107**(6-7): p. 507-513.

6. Palatinus, L., et al., Specifics of the data processing of precession electron diffraction tomography data and their implementation in the program PETS2. 0. Acta Crystallographica Section B: Structural Science, Crystal Engineering and Materials, 2019. **75**(4): p. 512-522.

7. Lanza, A., et al., Nanobeam precession-assisted 3D electron diffraction reveals a new polymorph of hen egg-white lysozyme. IUCrJ, 2019. **6**(2): p. 178-188.

8. Klar, P.B., et al., Accurate structure models and absolute configuration determination using dynamical effects in continuous-rotation 3D electron diffraction data. Nature Chemistry, 2023. **15**(6): p. 848-855.

9. Cichocka, M.O., Ångström, J., Wang, B., Zou, X., Smeets, S., High-throughput continuous rotation electron diffraction data acquisition via software automation. Applied Crystallography, 2018. **51**(6): p. 1652-1661.

10. Wang, B., X. Zou, and S. Smeets, Automated serial rotation electron diffraction combined with cluster analysis: an efficient multi-crystal workflow for structure determination. IUCrJ, 2019. **6**(5): p. 854-867.

11. Diouf, F. and M. Gemmi, LibraEDT: a software solution for automated 3D-ED data acquisition. Applied Crystallography, 2025. **58**(5): p. 1677 - 1687.

12. Gemmi, M., La Placa M.G.I., Galanis, A.S., Rauch, E.F., Nicolopoulos, S., Fast electron diffraction tomography. Applied Crystallography, 2015. **48**(3): p. 718-727.

13. Santucci, M. and U. Kolb, Evaluation of the reproducibility and crystal tracking precision of TEM goniometers in tomography experiments. Ultramicroscopy, 2026: p. 114308.

14. Leung, H.W., Copleym R.C.B., Laulainen, J.E.M., Johnstone, D.N., Midgley, P.A., Combining Scanning Nanobeam Electron Diffraction with 3D Electron Diffraction to Investigate Crystal Defects. 2024, Oxford University Press US.

15. Ophus, C., Four-Dimensional Scanning Transmission Electron Microscopy (4D-STEM): From Scanning Nanodiffraction to Ptychography and Beyond. Microscopy and Microanalysis, 2019. **25**(3): p. 563-582.

16. Eggeman, A.S., R. Krakow, and P.A. Midgley, Scanning precession electron tomography for three-dimensional nanoscale orientation imaging and crystallographic analysis. Nature communications, 2015. **6**(1): p. 7267.

17. Gallagher-Jones, M., et al., Atomic structures determined from digitally defined nanocrystalline regions. IUCrJ, 2020. **7**(3): p. 490-499.

18. Saha, A., et al., Reuniting crystallography with real space: Ab initio structure elucidation with 4D-STEM. Proceedings of the National Academy of Sciences, 2025. **122**(42): p. e2508185122.

19. Zhang, D., Oleynikov, P., Hovmöller, S., Zou, X., Collecting 3D electron diffraction data by the rotation method. Zeitschrift für Kristallographie, 2010. **225**(2-3): p. 94-102.

20. Midgley, P.A. and A.S. Eggeman, Precession electron diffraction–a topical review. IUCrJ, 2015. **2**(1): p. 126-136.

21. Martirosyan, M., et al., Nanoscale characterization of atomic positions in orthorhombic perovskite thin films. Small, 2025. **21**(43): p. e02538.

22. Passuti, S., et al., Scanning precession electron tomography (SPET) for structural analysis of thin films along their thickness. Symmetry, 2023. **15**(7): p. 1459.

23. Jannis, D., et al., Event driven 4D STEM acquisition with a Timepix3 detector: Microsecond dwell time and faster scans for high precision and low dose applications. Ultramicroscopy, 2022. **233**: p. 113423.

24. Annys, A., Lalandec Robert, H.L., Gholam, S., Hadermann, J., Verbeeck, J., Removing constraints of 4D-STEM with a framework for event-driven acquisition and processing. Ultramicroscopy, 2025. **277**: p. 114206.

25. Palatinus, L., V. Petříček, and C.A. Corrêa, Structure refinement using precession electron diffraction tomography and dynamical diffraction: theory and implementation. Foundations of Crystallography, 2015. **71**(2): p. 235-244.

26. Mugnaioli, E., T. Gorelik, and U. Kolb, "Ab initio" structure solution from electron diffraction data obtained by a combination of automated diffraction tomography and precession technique. Ultramicroscopy, 2009. **109**(6): p. 758-765.

27. Nederlof, I., van Genderen, E., Li, Y.W., Abrahams, J.P., A Medipix quantum area detector allows rotation electron diffraction data collection from submicrometre three-dimensional protein crystals. Biological Crystallography, 2013. **69**(7): p. 1223-1230.

28. Lukezic, A., Vojir, T., Zajc, L.C., Matas, J., Kristan M., Discriminative correlation filter with channel and spatial reliability. in Proceedings of the IEEE conference on computer vision and pattern recognition. 2017.

29. Bradski, G., The opencv library. Dr. Dobb's Journal: Software Tools for the Professional Programmer, 2000. **25**(11): p. 120-123.

30. Ravi, N., et al., Sam 2: Segment anything in images and videos. arXiv preprint arXiv:2408.00714, 2024.

31. Zhou, X.-G., et al., Probing the electron beam-induced structural evolution of halide perovskite thin films by scanning transmission electron microscopy. The Journal of Physical Chemistry C, 2021. **125**(19): p. 10786-10794.

32. Cleverley, A. and R. Beanland, Modelling fine-sliced three dimensional electron diffraction data with dynamical Bloch-wave simulations. IUCrJ, 2023. **10**(1): p. 118-130.

33. Palatinus, L. and G. Chapuis, SUPERFLIP–a computer program for the solution of crystal structures by charge flipping in arbitrary dimensions. Journal of Applied Crystallography, 2007. **40**(4): p. 786-790.

34. Sheldrick, G., SHELXT - Integrated space-group and crystal-structure determination. Acta Crystallographica Section A, 2015. **71**(1): p. 3-8.

35. Kleemiss, F., N. Peyerimhoff, and M. Bodensteiner, Refinement of X-ray and electron diffraction crystal structures using analytical Fourier transforms of Slater-type atomic wavefunctions in Olex2. Applied Crystallography, 2024. **57**(1): p. 161-174.

36. Petříček, V., Palatinus, L., Plášil, J., Dušek, M., Jana2020 – a new version of the crystallographic computing system Jana. Zeitschrift für Kristallographie - Crystalline Materials, 2023: p. 271 - 282.

37. López, C.A., et al., Crystal Structure Features of CsPbBr3 Perovskite Prepared by Mechanochemical Synthesis. ACS Omega, 2020. **5**(11): p. 5931-5938.

38. Denisov, N., A. Orekhov, and J. Verbeeck, Edge-Detected 4DSTEM-effective low-dose diffraction data acquisition method for nanopowder samples in an SEM instrument. The European Physical Journal Applied Physics, 2025. **100**: p. 5.

39. Williams, D.B. and C.B. Carter, The transmission electron microscope, in Transmission

electron microscopy: a textbook for materials science. 1996, Springer. p. 3-17, 215, 229, 274.

40. Rocklin, M. Dask: Parallel computation with blocked algorithms and task scheduling. in SciPy. 2015.

41. Peña, F.d., et al., hyperspy/hyperspy: v2. 1.0. Zenodo https://doi. org/10.5281/zenodo, 2024. **12724131**.

42. Otsu, N., A threshold selection method from gray-level histograms. Automatica, 1975. **11**(285-296): p. 23-27.

43. Li, C.H. and C. Lee, Minimum cross entropy thresholding. Pattern recognition, 1993. **26**(4): p. 617-625.

44. Zhang, Q., Diao, F., Cai, R., Zhou, G., Wang, Y., Atomic-Insights into electron-beam irradiation-induced degradation in all-inorganic metal halide perovskites. Acta Materialia, 2026. **309**: p. 122070.

45. Graafsma, H., et al., Detector developments for photon science at DESY. Frontiers in physics, 2024. **11**: p. 1321541.

46. Gholam, S. and J. Hadermann, The effect of the acceleration voltage on the quality of structure determination by 3D-electron diffraction. Ultramicroscopy, 2024. **266**.

47. Pratiush, U., Houston, A., Kalinin, S.V., Duscher, G., Realizing smart scanning transmission electron microscopy using high performance computing. Review of Scientific Instruments, 2024. **95**(10).

48. Sader, K., et al., Smart acquisition EELS. Ultramicroscopy, 2010. **110**(8): p. 998-1003.

49. Bergh, T., et al., Nanocrystal segmentation in scanning precession electron diffraction data. Journal of Microscopy, 2020. **279**(3): p. 158-167.

50. Francis, C. and P.M. Voyles, Clustering characteristic diffraction vectors in 4-D STEM data sets from overlapping structures in nanocrystalline and amorphous materials. Ultramicroscopy, 2024. **267**: p. 114040.

51. Martineau, B.H., Johnstone, D.N., van Helvoort, A.T.J., Midgley, P.A., Eggeman, A.S, Unsupervised machine learning applied to scanning precession electron diffraction data. Advanced structural and chemical imaging, 2019. **5**(1): p. 3.

52. Eremin, D.B., et al., Spatially Aware Diffraction Mapping Enables Fully Autonomous

MicroED. Journal of the American Chemical Society, 2025. **147**(46): p. 42299-42310.

53. Bera, S., R.K. Behera, and N. Pradhan, α-halo ketone for polyhedral perovskite nanocrystals: evolutions, shape conversions, ligand chemistry, and self-assembly. Journal of the American Chemical Society, 2020. **142**(49): p. 20865-20874.

54. Niermann, T. temscript: Python wrapper for the scripting interface of Thermo Fisher Scientific and FEI microscopes. 2022 [cited 2023; Available from: https://github.com/niermann/temscript/releases.

55. Brázda, P., Klementová, M., Krysiak, Y., Palatinus, L., Accurate lattice parameters from 3D electron diffraction data. I. Optical distortions. IUCrJ, 2022. **9**(6).

## 9. Supplementary Data

Supporting Information is available in the next page. The 3D ED datasets and CIFs are available in the Zenodo repository at 10.5281/zenodo.18939656.

## 10. Table of Contents

- 4D-STEM tomography is used as a new methodology for 3D ED data acquisition to solve structures.
- Object tracking and segmentation methods are applied for the extraction of 3D ED datasets from regions of interest in a 4D-STEM tomogram.
- Owing to its nanometric resolution, 4D-STEM tomography can tackle challenging samples being too small for 3D ED, have agglomerations or are beam-sensitive.

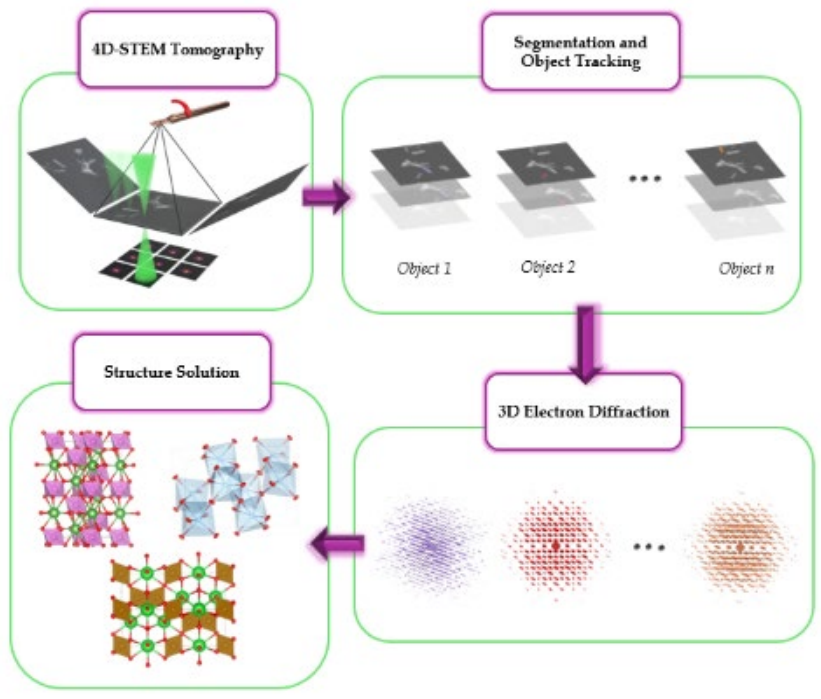

**Figure 16** Graphical abstract.

## 11. Supporting Information

### 11.1. R-factor Definitions

The R-factors referred in the paper are calculated based on the following formula.

$$R_{int} = \frac{\sum_{hkl}\sum_i |I_{hkl,i} - \langle I_{hkl}\rangle|}{\sum_{hkl}\sum_i I_{hkl,i}} \quad (1)$$

$$R_{meas} = \frac{\sum_{hkl}\left[\frac{n_{hkl}}{n_{hkl}-1}\right]^{\frac{1}{2}}\sum_i |I_{hkl,i} - \langle I_{hkl}\rangle}{\sum_{hkl}\sum_i I_{hkl,i}} \quad (2)$$

$$R_{obs} = \frac{\sum_{hkl}\big||F_o| - |F_c|\big|}{\sum_{hkl}|F_o|} \quad (3)$$

In these equations, $I_{hkl,i}$ is the intensity of $i$-th observation for reflection $hkl$, $\langle I_{hkl}\rangle$ is the mean intensity of all symmetry-equivalent observations for reflection $hkl$, $n_{hkl}$ number of symmetry-equivalent observations of reflection $hkl$, and $F_o$ and $F_c$ are respectively observed and calculated structure factor amplitudes.

### 11.2. Threshold for Skipping Reflections

Omitting reflections is a sensitive subject in crystallographic studies for structure solution, as this can force the model toward a specific direction against observations. However, coincidence losses and localized pixel saturations occur on detectors with Timepix3 chips at high fluxes.[1] These introduce direct errors on the integration process. This is highly likely in parallel beam illumination as the collected signal is localized at Bragg spots. Although it is attempted to reject these errors for the kinematical refinement using error-model adjustments implemented in PETS2 software [2], the dynamical refinement uses intensities of reflections at each frame to form virtual frames [3], and consequently such counting errors will be included. Omission of the reflections are performed here when the model is refined and stabilized after an initial dynamical refinement. It is essential to note that the distinction of the clipping effect is not trivial as it does not occur at a fixed value for the event-based detectors. Additionally, it is even more complex to measure this for the extracted frames from 4D-STEM tomography, since frames are the result of summing a varying region on 4D-STEM tomography data and at different crystal orientations.

To select an effective and yet a safe threshold for omitting reflections during dynamical refinement, a range of values were checked during the dynamical refinement of T-P1 and T-P2 particles. In Jana software, the threshold is based on the following equation:

$$|F_{obs} - F_{calc}| > threshold \times sig(F_{obs}) \quad (4)$$

**Figure S 1** demonstrates the results after tweaking this threshold in a wide range and its influence on the number of reflections which are skipped, dynamical refinement $R_{obs}$ and the largest residual charge in the difference Fourier maps. Both particles show a sudden decrease in $R_{obs}$ in the range between 20 and 30. A closer look at the reflection profiles and 3D ED frames showed most of the reflections culled in this range are strong reflections close to a zone axis, and some of them even have saturated intensities in their intensity profile. Thus, it is highly likely that these reflections suffer from both the detector non-linearities and heavy dynamical effects. The convolution of these effects can cause deviation in the dynamical refinement. Threshold 25 was selected as a more conservative threshold for 3D ED datasets extracted from DS-1 dataset. The threshold was lowered to 20 for DS-3 dataset with $CsPbBr_3$.

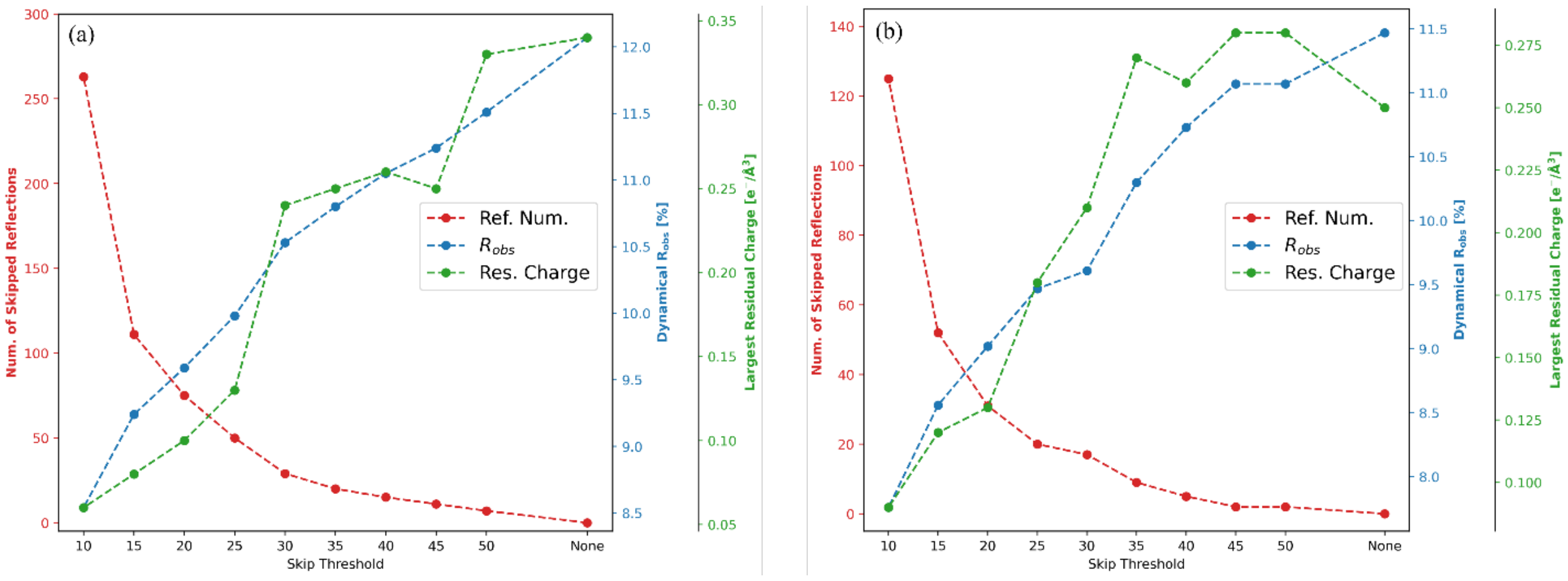

**Figure S 1**. The influence of the threshold for skipping reflections on the dynamical refinement of the brookite structure; (a) T-P1, (b) T-P2

### 11.3. Probe Size in STEM

The achievable probe size can be estimated by the squared contribution of the beam spreading effects. [4, 5] Assuming Gaussian shape and independence of the these effects, the probe size is influenced by the contributions from the initial Gaussian diameter at the gun ($d_g$), diffraction at the final aperture ($d_d$), the spherical aberration effects in the beam-forming lens ($d_s$), and the chromatic aberration blurring ($d_c$). For a TEM without aberration-correction, these can be estimated as:

$$d_g = \frac{2}{\pi\alpha}\sqrt{\frac{I_p}{\beta}} \quad (5)$$

$$d_d = \frac{0.61\lambda}{\alpha} \quad (6)$$

$$d_s = 0.5C_s\alpha^3 \quad (7)$$

$$d_c = C_c\left(\frac{\Delta E}{E_0}\right) \quad (8)$$

In these equations, $\alpha$, $\beta$, and $I_p$ respectively stand for the convergence angle, the brightness of the source and the probe current, $\lambda$ is the wavelength of the accelerated electrons, $C_s$ is the spherical aberration coefficient, and $C_c$, $\Delta E$ and $E_0$ are respectively the chromatic aberration coefficient, the energy spread of the electron beam and the energy of the electrons. Then, the final probe size can be calculated as:

$$d_{probe} = \sqrt{d_g^2 + d_d^2 + d_s^2 + d_c^2} \quad (9)$$

The assumptions for the calculation of the probe size are listed in **Table S 1**. For achieving the probe FWHM according to Kirkland [6], the prefactors of the abovementioned components should be respectively changed to 0.514, 0.18, 0.6 and 0.707.

**Table S 1**. The estimated microscope parameters used for the theoretical probe size calculations.

| $\boldsymbol{\beta}\ [A.m^{-2}sr^{-1}]$ | $\boldsymbol{I_p}\ [A]$ | $\boldsymbol{\lambda}\ [pm]$ | $\boldsymbol{C_s}\ [mm]$ | $\boldsymbol{C_c}\ [mm]$ | $\boldsymbol{\Delta E}\ [eV]$ | $\boldsymbol{E_0}\ [eV]$ |
|---|---|---|---|---|---|---|
| $\mathbf{5\times10^{12}}$ | $5\times10^{-12}$ | 2.508 | 1.5 | 1.5 | 0.5 | 200,000 |

### 11.4. Disk Overlaps in Diffraction

As the convergence of the electron beam increases, the size of the reflections increases in the reciprocal space. To be able to follow 3D ED data analysis, the disk overlaps should be avoided. On one hand, there is a relation between the size of the disks and the convergence angle. On the other hand, the distance between the disks depends on the size of the unit cell and the orientation of the crystal. Thus, there is a relation between the size of the unit cell and the largest convergence angle which can be selected. As illustrated in **Figure S2(a)**, by trigonometry $D_{disk} = 2\alpha/\lambda$. In the reciprocal space, the minimum distance between the disks are when the crystal orients along the maximum interplanar spacing (d-spacing) of the structure. Therefore, the condition $D_{disk} < \frac{1}{d\text{-}spacing^{max}}$ is the key for avoiding disk overlaps. Thus, for studying a structure, d-spacing should be smaller than $\lambda/2\alpha$. For an orthorhombic structure (such as the samples studied in this work), largest d-spacing equals to the maximum of the unit cell parameters a, b and c. With respect to the probable charge spreading of the incoming electrons in the detector, it would be safer to keep the convergence somewhat smaller than such ideal calculation so that there would be no chance of disk overlaps. It should be noted that these limitations are relaxed at lower acceleration voltages as they have higher wavelengths (**Figure S 2b**).

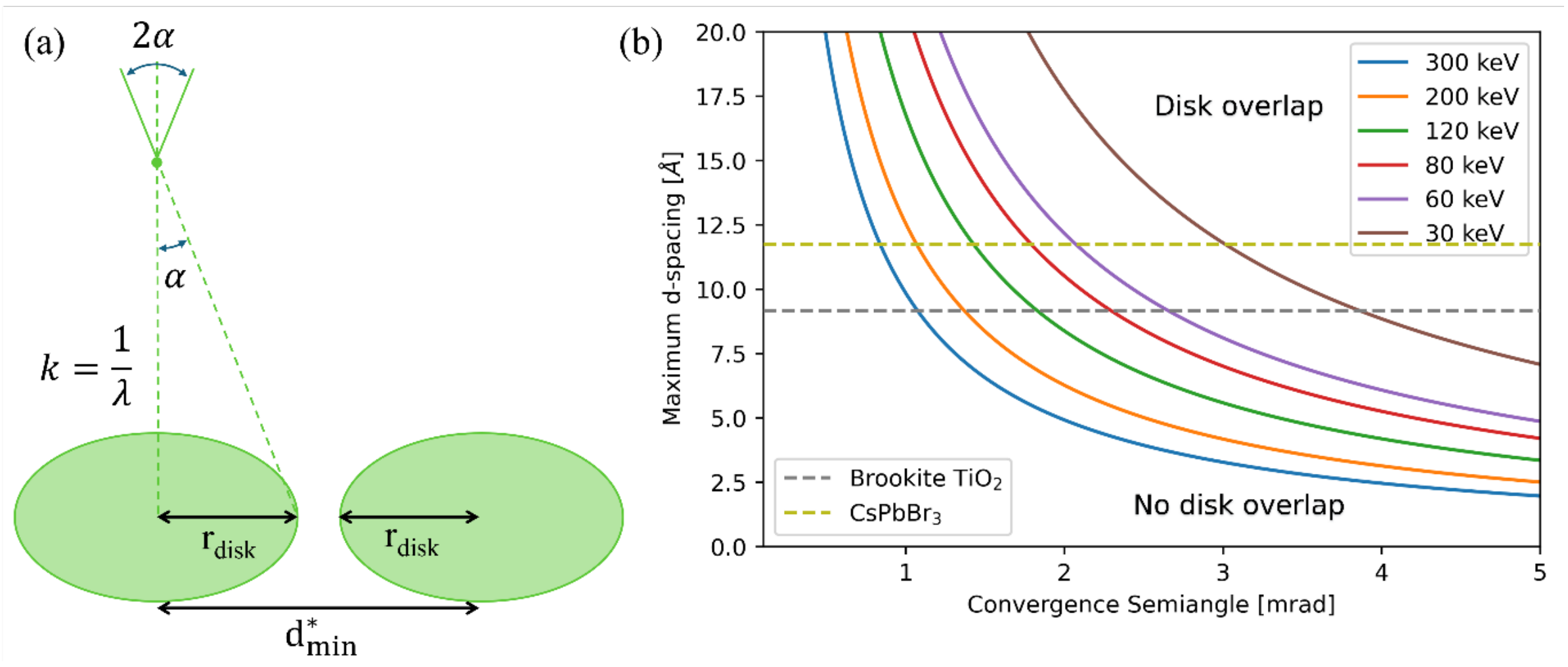

**Figure S 2**. (a) The relationship between the convergence angle and the overlap of the reflection disks; (b) the limits of d-spacing for not having disk overlap at different acceleration voltages.

## 11.5. Supplementary Figures

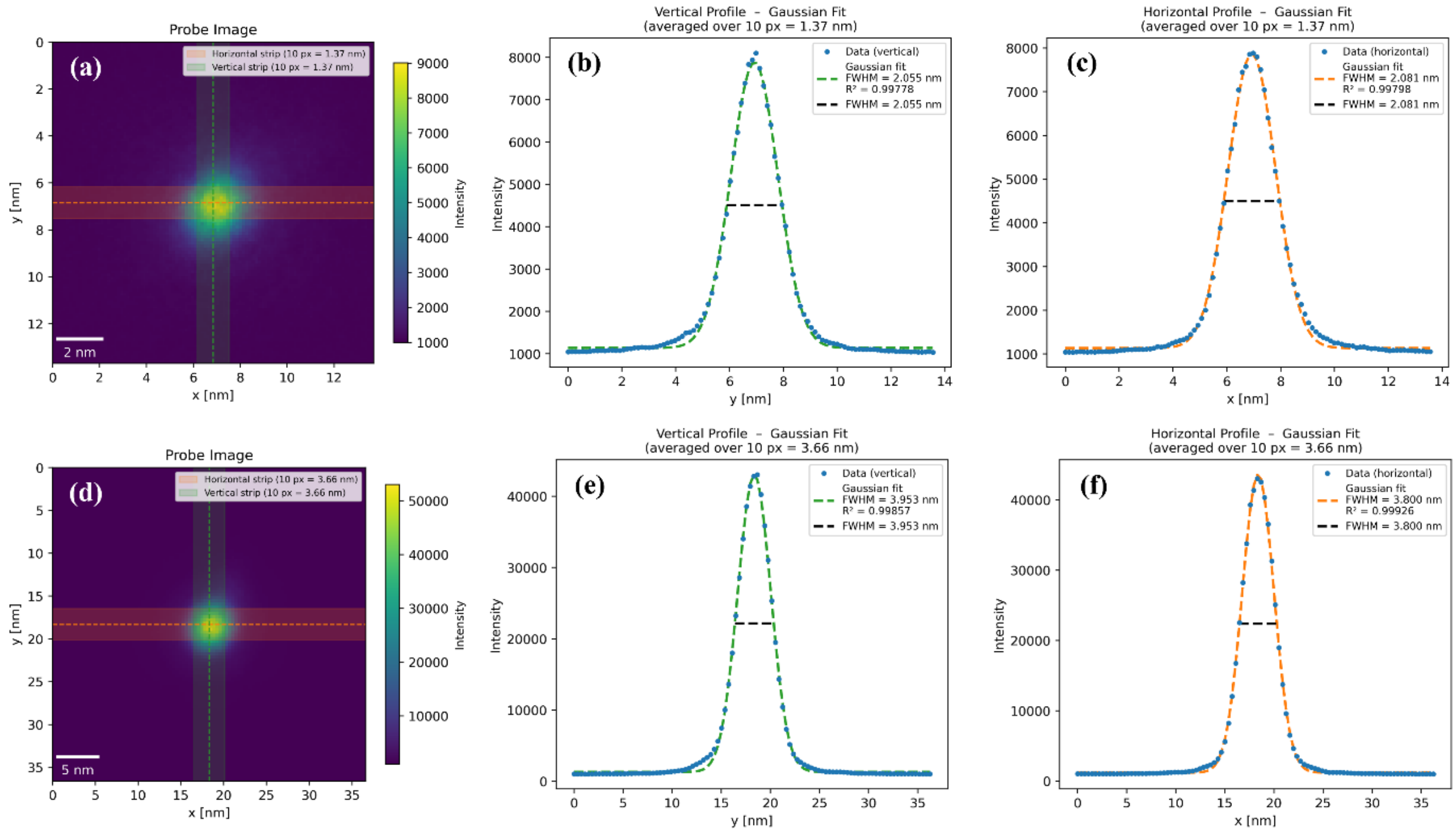

**Figure S 3**. The probe image formed using (a) 20 µm aperture and (b) 10 µm aperture with their Gaussian fit for FWHM calculations in (b, e) vertical and (c, f) horizontal directions. The average of the horizontal and vertical values are reported in the main text. The images were acquired by a US1000 CCD detector (Gatan).

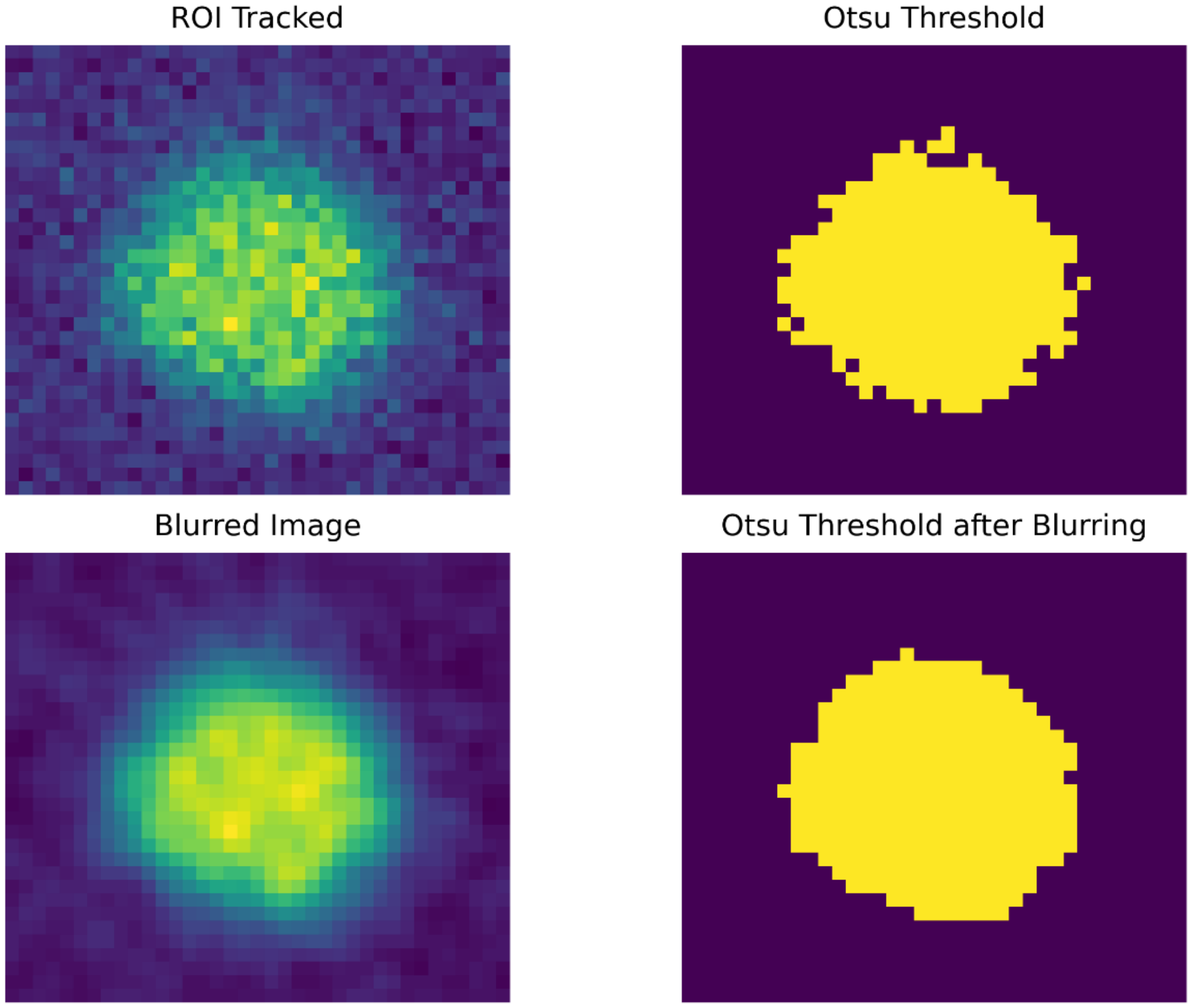

**Figure S 4**. An example of blurring and binarization on a tracked ROI on C-P2.

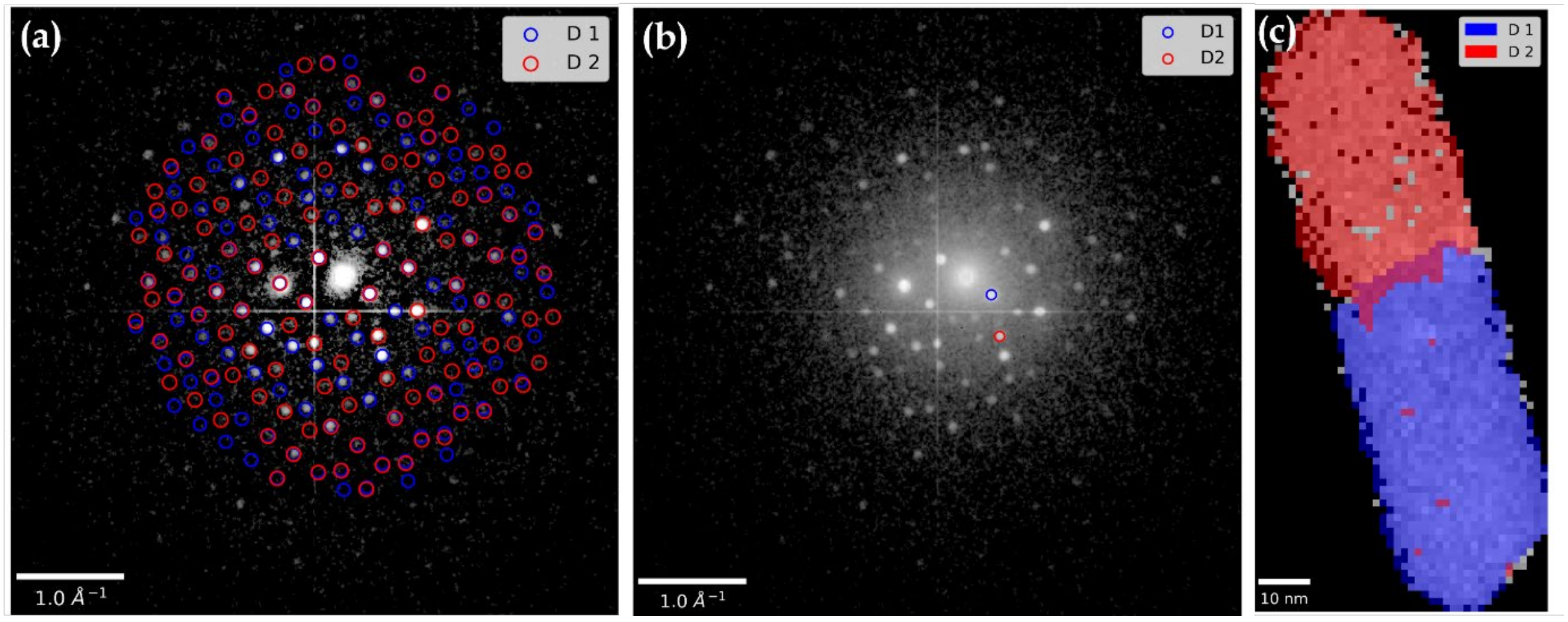

**Figure S 5**. The presence of two different domains in T-P5, i.e. D1 and D2; (a) overlay of the integration masks for D1 and D2 in the 3D ED dataset, (b) selected reflections as virtual detectors for visualizing D1 and D2, (c) the overlay of the domain masks over the particle.

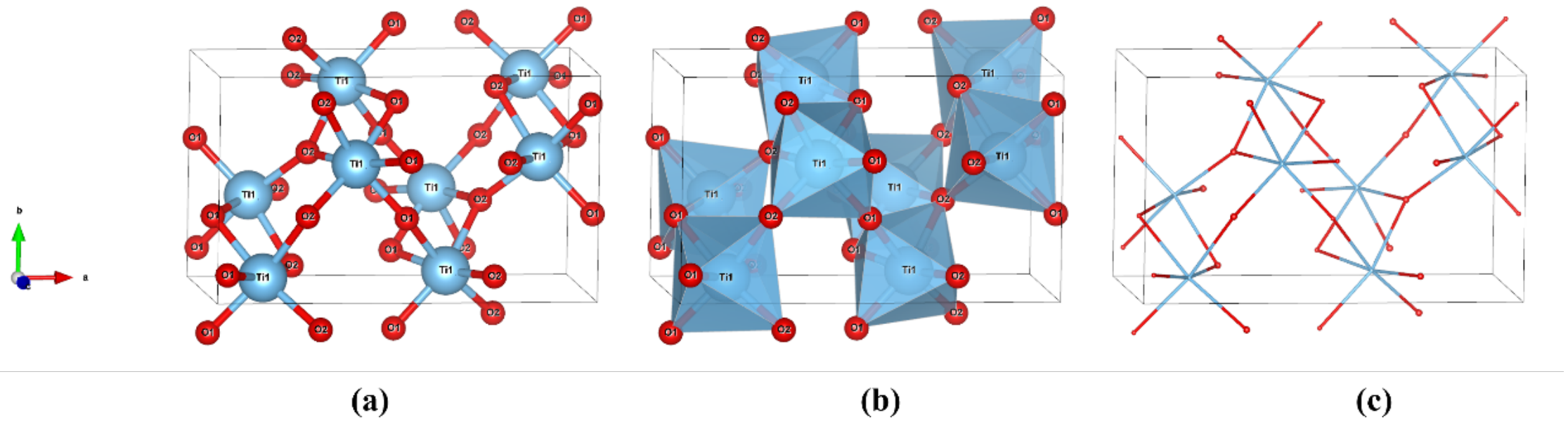

**Figure S 6**. Brookite atomic structure in (a) ball and stick model, (b) polyhedral model and (c) with 50% probability for atomic movements. The structure is based on T-P4 after dynamical refinement with culled reflections.

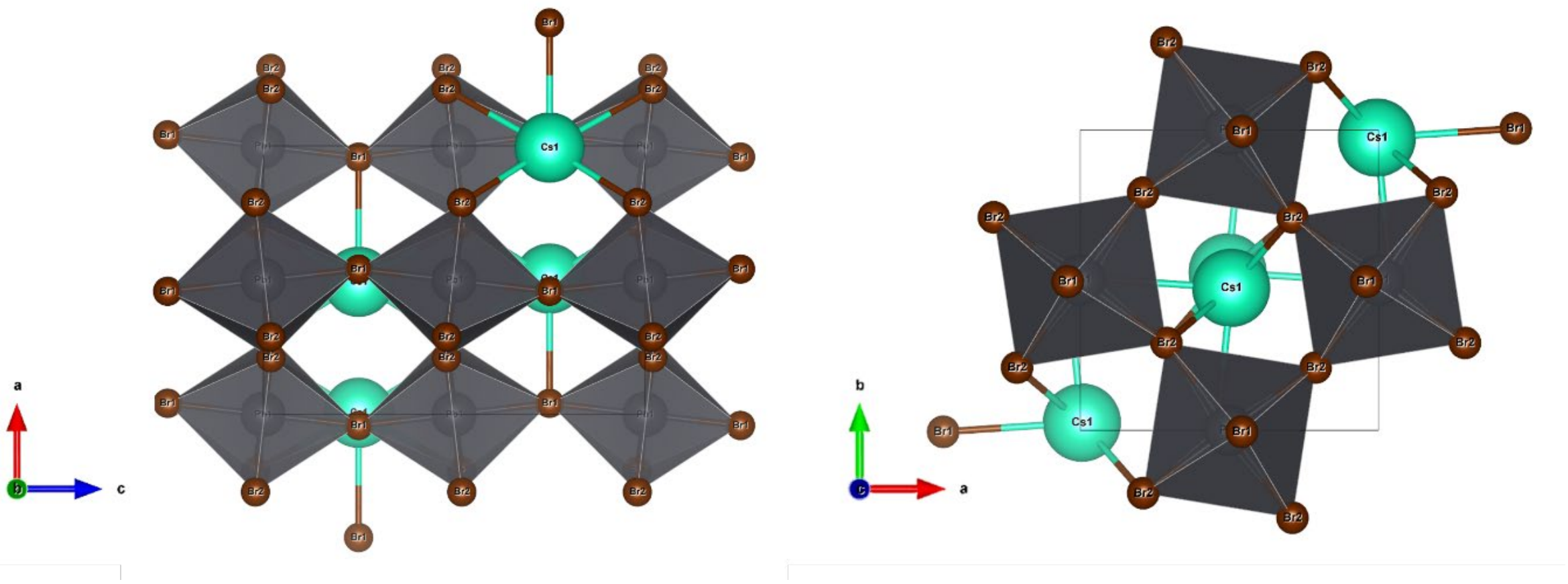

**Figure S 7**. $CsPbBr_3$ atomic structure from different view axes. The tilted polyhedra can be observed in both views. The model is based on C-P1 after dynamical refinement with culled reflections.

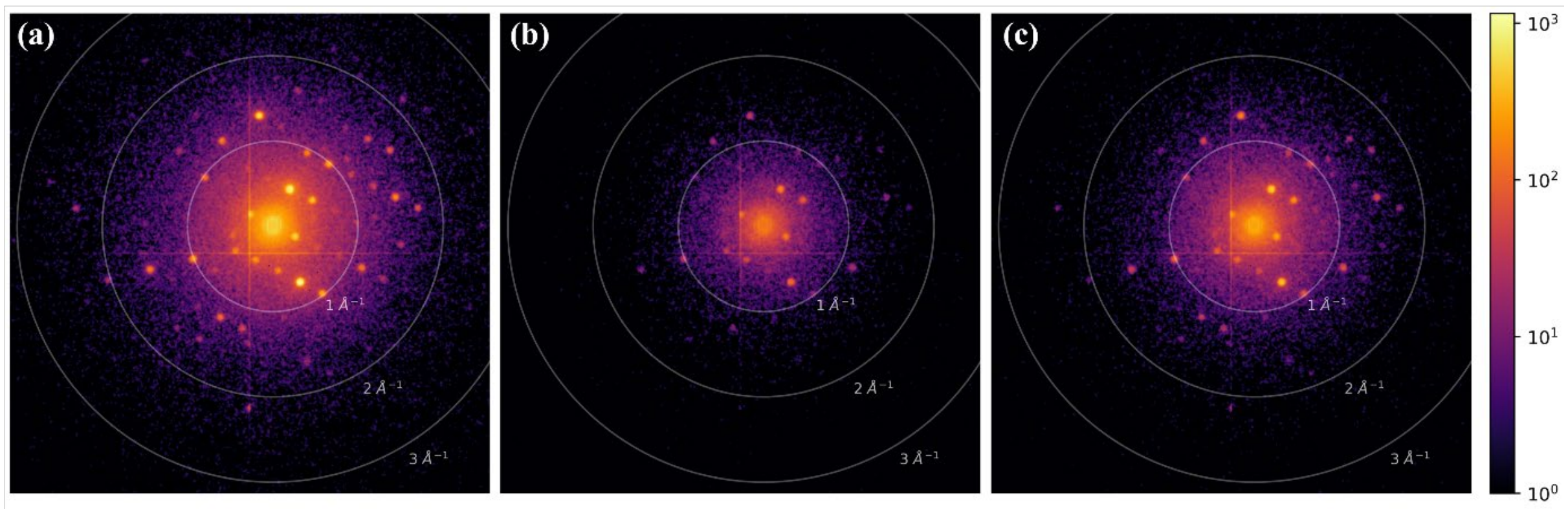

**Figure S 8**. Diffraction patterns at α = -50° for (a) T-P1, (b) T-P1-E1 and (c) T-P1-E2.

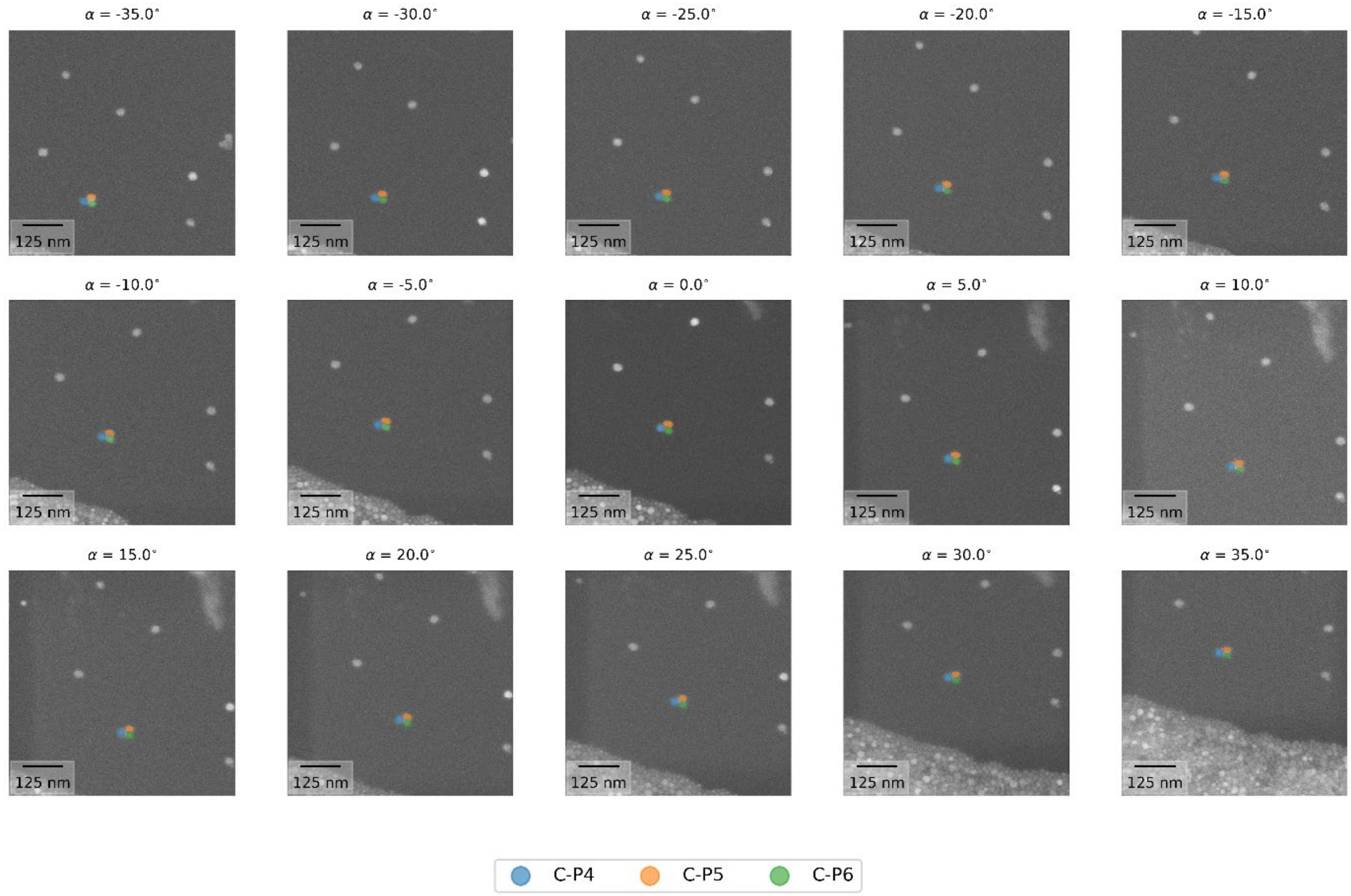

**Figure S 9.** Segmentation and tracking using SAM2 tracker on the agglomerate of some particles in DS-3 dataset.

### 11.6. Supplementary Tables

In the upcoming tables, the following terms are used:

- Nall: Total number of unique reflections
- Nobs: Number of unique, observed reflections
- Completeness: the percentage of observed reflections by considering the symmetry of the system, showing the coverage of the reciprocal space
- Redundancy: The average number of times a reflection is observed in the dataset
- <I/σ>: The average signal-to-noise ration of the measured intensities
- wR(obs): Weighted R-factors during refinement calculated on observed reflections

**Table S 2**. Data reduction statistics for 3D-ED datasets extracted from $TiO_2$ particles in dataset DS-1.

| | | *T-P1* | *T-P2* | *T-P3* | *T-P4* | *T-P1-E1* | *T-P1-E2* |
|---|---|---|---|---|---|---|---|
| *Tilt Range [°]* | | [-50, 30] | [-50, 28] | [-50, 30] | [-50, 30] | [-50, 30] | [-50, 30] |
| *Res. [Å$^{-1}$]* | | 2.0 | 2.0 | 2.0 | 2.0 | 2.0 | 2.0 |
| *Found Unit Cell*[a] | a [Å], | 9.157(2) | 9.144(3) | 9.200(1) | 9.159(3) | 9.178(7) | 9.168(4) |
| | b [Å], | 5.469(0) | 5.397(1) | 5.445(1) | 5.433(1) | 5.471(1) | 5.469(1) |
| | c [Å] | 5.075(0) | 5.157(0) | 5.075(0) | 5.116(0) | 5.073(1) | 5.072(0) |
| | α [°], | 90.06(1) | 89.43(1) | 89.79(0) | 90.87(1) | 90.00(2) | 89.96(1) |
| | β [°], | 89.80(1) | 89.85(1) | 89.94(1) | 89.50(1) | 89.70(3) | 89.77(1) |
| | γ [°] | 90.29(1) | 90.47(2) | 90.35(1) | 89.93(3) | 90.19(4) | 89.77(2) |
| *Nobs / Nall* | | 269 / 921 | 504 / 1014 | 298 / 998 | 583 / 1013 | 240 / 927 | 293 / 922 |
| *Redundancy* | | 4.16 | 3.64 | 3.80 | 3.77 | 4.14 | 4.15 |
| *Completeness [%]* | | 71.8 | 79.1 | 77.9 | 79.0 | 72.3 | 71.9 |
| *<I/σ>* | | 3.45 | 4.31 | 3.79 | 4.93 | 3.48 | 3.52 |
| *$R_{int}$ (obs) [%]* | | 16.37 | 10.85 | 12.71 | 9.10 | 11.17 | 14.25 |
| *$R_{meas}$ (obs) [%]* | | 18.94 | 12.80 | 14.90 | 10.54 | 12.99 | 16.44 |

[a] The reported unit cell is the raw cell found in the data. Then, the distortion refinements were performed using the fixed cell method reported in the previous work based on synchrotron XRD results. [6]

[b] As the data is step-wise, and not continuous, PETS2 calculates high partialities with respect to the lost data between the collected points.

**Table S 3**. Structure refinement statistics for the $TiO_2$ particles in dataset DS-1.

| **Type** | **Parameter** | **T-P1** | **T-P2** | **T-P3** | **T-P4** | **T-P1-E1** | **T-P1-E2** |
|---|---|---|---|---|---|---|---|
| Kinematical | # Ref. / # Param. | 591 / 13 | 725 / 13 | 562 / 13 | 755 / 13 | 506 / 13 | 599 / 13 |
| | Goodness of Fit | 3.81 | 3.61 | 3.02 | 4.09 | 2.48 | 3.23 |
| | R(obs) [%] | 26.28 | 21.50 | 19.35 | 20.46 | 16.42 | 21.77 |
| | wR(obs) [%] | 30.22 | 25.64 | 23.58 | 25.68 | 20.06 | 25.91 |
| | Avg. Distance [Å] | 0.018 | 0.014 | 0.015 | 0.012 | 0.013 | 0.016 |
| | Max Distance [Å] | 0.027 | 0.026 | 0.028 | 0.018 | 0.024 | 0.026 |
| Kinematical with Extinction Correction | # Ref. / # Param. | - | 725 / 14 | 562 / 14 | - | 506 / 14 | 599 / 14 |
| | Goodness of Fit | - | 3.37 | 2.57 | - | 2.39 | 2.92 |
| | R(obs) [%] | - | 18.40 | 13.65 | - | 14.22 | 18.76 |
| | wR(obs) [%] | - | 23.94 | 20.06 | - | 19.34 | 23.42 |
| | Avg. Distance [Å] | - | 0.013 | 0.014 | - | 0.013 | 0.016 |
| | Max Distance [Å] | - | 0.026 | 0.020 | - | 0.023 | 0.025 |
| Dynamical | # Ref. / # Param. | 3243 / 56 | 2686 / 55 | 2517 / 56 | 2640 / 56 | 1697 / 56 | 2683 / 56 |
| | Goodness of Fit | 7.12 | 5.65 | 5.19 | 5.81 | 2.91 | 4.86 |
| | R(obs) [%] | 12.15 | 11.69 | 11.44 | 12.68 | 9.97 | 10.92 |
| | wR(obs) [%] | 14.30 | 13.73 | 13.53 | 14.69 | 10.41 | 12.76 |
| | Avg. Distance [Å] | 0.006 | 0.005 | 0.007 | 0.005 | 0.010 | 0.008 |
| | Max Distance [Å] | 0.009 | 0.006 | 0.008 | 0.006 | 0.018 | 0.015 |
| Dynamical with Skipping Reflections | # Ref. / # Param. | 3194 / 56 | 2686 / 55 | 2493 / 56 | 2614 / 56 | 1697 / 56 | 2667 / 56 |
| | # Culled Ref. | 49 | 21 | 24 | 26 | 0 | 4.05 |
| | Goodness of Fit | 5.33 | 4.22 | 3.93 | 4.66 | 2.91 | 16 |
| | R(obs) [%] | 9.63 | 9.41 | 9.32 | 11.16 | 9.97 | 9.63 |
| | wR(obs) [%] | 10.91 | 10.42 | 10.46 | 12.06 | 10.41 | 10.78 |
| | Avg. Distance [Å] | 0.006 | 0.007 | 0.008 | 0.006 | 0.010 | 0.008 |
| | Max Distance [Å] | 0.011 | 0.010 | 0.010 | 0.008 | 0.018 | 0.015 |
| Dynamical with Anisotropic ADP | # Ref. / # Param. | 3194 / 71 | 2665 / 70 | 2494 / 71 | 2614 / 71 | 1697 / 71 | 2666 / 71 |
| | # Culled Ref. | 49 | 21 | 23 | 26 | 0 | 17 |
| | Goodness of Fit | 5.22 | 4.15 | 3.92 | 4.62 | 2.88 | 3.97 |
| | R(obs) [%] | 9.42 | 9.24 | 9.22 | 11.02 | 9.85 | 9.41 |
| | wR(obs) [%] | 10.66 | 10.23 | 10.41 | 11.93 | 10.23 | 10.54 |
| | Avg. Distance [Å] | 0.006 | 0.006 | 0.008 | 0.006 | 0.010 | 0.008 |
| | Max Distance [Å] | 0.010 | 0.009 | 0.009 | 0.007 | 0.014 | 0.014 |

**Table S 4**. Bond Valence Sum (BVS) values of $TiO_2$ atoms for the solved structures in datasets DS-1. The BVS of the reference structure (ICSD ID 154605) was 4.025(17), 2.074(13) and 1.951(11) for respectively Ti 1, O 1 and O2 atoms.

| | | T-P1 | T-P2 | T-P3 | T-P4 | T-P1-E1 | T-P1-E2 |
|---|---|---|---|---|---|---|---|
| **Kinematical** | Ti 1 | 4.02(3) | 4.03(2) | 4.01(3) | 4.02(2) | 4.02(3) | 4.02(3) |
| | O 1 | 2.07(2) | 1.97(2) | 2.04(2) | 2.07(2) | 2.06(2) | 2.06(2) |
| | O 2 | 1.96(2) | 2.06(2) | 1.97(2) | 1.96(1) | 1.96(2) | 1.96(2) |
| **Kinematical with Extinction Correction** | Ti 1 | - | 4.03(2) | 4.01(2) | - | 4.02(3) | 4.02(2) |
| | O 1 | - | 1.95(1) | 2.05(2) | - | 2.06(2) | 2.07(2) |
| | O 2 | - | 2.08(2) | 1.96(2) | - | 1.96(2) | 1.96(2) |
| **Dynamical** | Ti 1 | 4.02(1) | 4.02(1) | 4.02(1) | 4.02(1) | 4.02(2) | 4.02(1) |
| | O 1 | 2.058(8) | 1.958(9) | 2.060(9) | 2.07(1) | 2.05(1) | 2.056(9) |
| | O 2 | 1.959(7) | 2.06(1) | 1.958(8) | 1.953(9) | 1.97(1) | 1.960(8) |
| **Dynamical with Skipped Reflections** | Ti 1 | 4.020(7) | 4.025(9) | 4.020(9) | 4.02(1) | 4.02(2) | 4.018(8) |
| | O 1 | 2.058(5) | 1.955(6) | 2.064(7) | 2.070(7) | 2.06(1) | 2.068(6) |
| | O 2 | 1.961(5) | 2.070(7) | 1.956(6) | 1.950(6) | 1.97(1) | 1.950(6) |
| **Dynamical with Anisotropic ADP** | Ti 1 | 4.019(7) | 4.026(9) | 4.020(9) | 4.02(1) | 4.02(2) | 4.018(8) |
| | O 1 | 2.058(5) | 1.956(6) | 2.064(7) | 2.070(7) | 2.06(1) | 2.068(6) |
| | O 2 | 1.961(5) | 2.070(7) | 1.956(6) | 1.951(6) | 1.97(1) | 1.950(5) |

**Table S 5**. Data reduction statistics for 3D-ED datasets extracted from $CsPbBr_3$ particles in dataset DS-3.

| **Particle Code** | | **C-P1** | **C-P2** | **C-P3** |
|---|---|---|---|---|
| **Tilt Range [°]** | | [-35, 34] | [-35, 35] | [-35, 35] |
| **Resolution [Å$^{-1}$]** | | 1.4 | 1.4 | 1.4 |
| **Found Unit Cell** | a [Å] | 8.296(2) | 8.270(1) | 8.239(3) |
| | b [Å] | 8.357(2) | 8.258(2) | 8.353(4) |
| | c [Å] | 11.738(3) | 11.846(5) | 11.788(7) |
| | α [°] | 90.18(2) | 90.40(4) | 89.03(6) |
| | β [°] | 90.51(0) | 91.00(3) | 89.55(6) |
| | γ [°] | 90.86(0) | 89.34(3) | 89.78(4) |
| **Refined Unit Cell[a]** | a [Å] | 8.290(3) | 8.280(1) | 8.323(4) |
| | b [Å] | 8.326(3) | 8.328(1) | 8.304(4) |
| | c [Å] | 11.819(4) | 11.861(2) | 11.831(5) |
| | α [°] | 90(0) | 90(0) | 90(0) |
| | β [°] | 90(0) | 90(0) | 90(0) |
| | γ [°] | 90(0) | 90(0) | 90(0) |
| **Nobs / Nall** | | 379 / 1166 | 340 / 1056 | 364 / 1160 |
| **Redundancy** | | 2.94 | 3.28 | 2.98 |
| **Completeness [%]** | | 84.1 | 75.9 | 83.5 |
| **<I/σ>** | | 4.29 | 4.23 | 4.30 |
| **$R_{int}$ (obs) [%]** | | 9.35 | 8.97 | 10.63 |
| **$R_{meas}$ (obs) [%]** | | 11.44 | 10.83 | 12.82 |

[a] The refined unit cells are after distortion refinement by assuming orthorhombic symmetry.

**Table S 6**. Structure refinement statistics for the $CsPbBr_3$ particles in dataset DS-3.

| Refinement Type | Parameter | C-P1 | C-P2 | C-P3 |
|---|---|---|---|---|
| Kinematical | # Reflection / # Parameter | 530 / 12 | 502 / 12 | 495 / 12 |
| | Goodness of Fit | 3.10 | 3.59 | 3.86 |
| | R(obs) [%] | 17.87 | 18.75 | 19.64 |
| | wR(obs) [%] | 24.16 | 27.31 | 28.55 |
| Kinematical with Extinction Correction | # Reflection / # Parameter | 530 / 13 | 502 / 13 | 495 / 13 |
| | Goodness of Fit | 3.09 | 3.59 | 3.86 |
| | R(obs) [%] | 17.87 | 18.67 | 19.67 |
| | wR(obs) [%] | 24.11 | 27.30 | 28.55 |
| Dynamical | # Reflection / # Parameter | 1007 / 50 | 1240 / 51 | 962 / 51 |
| | Goodness of Fit | 4.00 | 3.28 | 3.68 |
| | R(obs) [%] | 12.83 | 10.64 | 12.18 |
| | wR(obs) [%] | 14.46 | 11.57 | 12.98 |
| Dynamical with Skipping Reflections | # Reflection / # Parameter | 1002 / 50 | 1240 / 51 | 961 / 51 |
| | # Culled Reflection | 5 | 0 | 1 |
| | Goodness of Fit | 3.64 | 3.28 | 3.59 |
| | R(obs) [%] | 12.34 | 10.64 | 12.04 |
| | wR(obs) [%] | 13.27 | 11.57 | 12.65 |
| Dynamical with Anisotropic ADP | # Reflection / # Parameter | 1003 / 66 | 1240 / 67 | 961 / 67 |
| | # Culled Reflection | 4 | 0 | 1 |
| | Goodness of Fit | 2.90 | 2.64 | 2.79 |
| | R(obs) [%] | 9.40 | 8.13 | 9.09 |
| | wR(obs) [%] | 10.43 | 9.26 | 9.75 |

**Table S 7**. Bond Valence Sum (BVS) values of $CsPbBr_3$ atoms for the solved structures in datasets DS-3.

| Refinement | Atom | C-P1 | C-P2 | C-P3 |
|---|---|---|---|---|
| **Kinematical** | Pb 1 | 2.38(2) | 2.36(2) | 2.42(3) |
| | Cs 1 | 0.63(1) | 0.65(1) | 0.62(1) |
| | Br 1 | 1.00(1) | 0.98(1) | 1.02(2) |
| | Br 2 | 1.00(1) | 1.01(2) | 1.01(1) |
| **Kinematical with Extinction Correction** | Pb 1 | 2.39(2) | 2.36(2) | 2.42(3) |
| | Cs 1 | 0.63(1) | 0.65(1) | 0.66(1) |
| | Br 1 | 1.00(1) | 0.98(1) | 1.02(2) |
| | Br 2 | 1.00(1) | 1.01(2) | 1.04(1) |
| **Dynamical** | Pb 1 | 2.39(2) | 2.35(1) | 2.41(2) |
| | Cs 1 | 0.63(1) | 0.63(0) | 0.60(1) |
| | Br 1 | 1.00(1) | 0.98(0) | 1.01(1) |
| | Br 2 | 1.00(1) | 1..00(1) | 1.00(1) |
| **Dynamical with Skipped Reflections** | Pb 1 | 2.39(2) | 2.35(1) | 2.42(1) |
| | Cs 1 | 0.63(1) | 0.63(0) | 0.59(1) |
| | Br 1 | 1.00(1) | 0.98(0) | 1.01(1) |
| | Br 2 | 1.00(1) | 1.00(1) | 1.00(1) |
| **Dynamical with Anisotropic ADP** | Pb 1 | 2.38(1) | 2.35(1) | 2.42(1) |
| | Cs 1 | 0.63(1) | 0.63(4) | 0.59(1) |
| | Br 1 | 1.00(1) | 0.98(0) | 1.01(1) |
| | Br 2 | 1.00(1) | 1.00(1) | 0.99(1) |

## 11.7. Supporting Information References

1. Annys, A., et al., Removing constraints of 4D-STEM with a framework for event-driven acquisition and processing. Ultramicroscopy, 2025. **277**: p. 114206.

2. Khouchen, M., et al., Optimal estimated standard uncertainties of reflection intensities for kinematical refinement from 3D electron diffraction data. Acta Crystallographica Section A, 2023. **79**(5): p. 427-439.

3. Klar, P.B., et al., Accurate structure models and absolute configuration determination using dynamical effects in continuous-rotation 3D electron diffraction data. Nature Chemistry, 2023. **15**(6): p. 848-855.

4. Kohl, H. and L. Reimer, Transmission Electron Microscopy. Springer Series in Optical Sciences. Vol. 36. 2008.

5. Williams, D.B. and C.B. Carter, The transmission electron microscope, in Transmission electron microscopy: a textbook for materials science. 1996, Springer. p. 3-17, 215, 274.

6. Kirkland, E.J., Advanced computing in electron microscopy. 3 ed. Vol. 12. 2020: Springer.